# Migration of a surfactant-laden droplet in non-isothermal Poiseuille flow

Sayan Das, Shubhadeep Mandal, S K Som and Suman Chakraborty*

*Department of Mechanical Engineering, Indian Institute of Technology Kharagpur, Kharagpur – 721302, India*

The motion of a surfactant-laden viscous droplet in the presence of background non-isothermal Poiseuille flow is studied analytically and numerically. Specifically, the effect of interfacial Marangoni stress due to non-uniform distribution of surfactants and temperature at the droplet interface on the velocity and direction of motion of the droplet along the centerline of imposed Poiseuille flow is investigated in the presence of linearly varying temperature field. In the absence of thermal convection, fluid inertia and shape deformation, the interfacial transport of bulk-insoluble surfactants is governed by the surface Péclet number which represents the relative strength of the advective transport of surfactant over the diffusive transport. We obtain analytical solution for small and large values of the surface Péclet number. Numerical solution is obtained for arbitrary surface Péclet number, which compares well with the analytical solution. Depending on the direction of temperature gradient with respect to the imposed Poiseuille flow, the surfactant-induced Marangoni stress affects the droplet velocity differently. When the imposed temperature increases in the direction of imposed Poiseuille flow, surfactants retard the droplet motion as compared with a surfactant-free droplet. However, when the imposed temperature decreases in the direction of imposed Poiseuille flow, presence of surfactants may increase or decrease the magnitude of droplet velocity depending on the relevant governing parameters. Further, for particular values of governing parameters, we observe change in direction of droplet motion due to presence of surfactants, which may bear significant consequences in the design of droplet based microfluidic systems.

## I. INTRODUCTION

The study of motion of droplets and bubbles in another immiscible carrier liquid medium is of utmost importance due a wide variety of applications, primarily in the bioengineering and biomedical scenario.[1–3] With the advent of novel emulsification techniques in microfluidic devices,[2,4] droplets are generated with unprecedented throughput and being used for drug delivery, protein crystallization, biomolecule synthesis, chemical reactions, nanoparticle synthesis, and single cell analysis.[5–9] Optimum functionalities of these processes in respective droplet-based devices are not only governed by the effective generation of droplets but also on the active control over the motion and pathway of droplets from one point to the other.[10,11]

*Corresponding author, email: suman@mech.iitkgp.ernet.in

Droplets are often transported through microchannles by applying pressure gradient. Several theoretical and experimental studies have been reported in the literature which considered the motion of droplets in Poiseuille flow.[12,13] A Newtonian liquid droplet of clean fluid-fluid interface (i.e., no surfactants), far away from the bounding walls, moves only in the axial direction in the absence of shape deformation and fluid inertia. In the presence of non-linear effects (e.g., shape deformation, fluid inertia and viscoelastic fluid rheology), the droplet located at off-centerline position can migrate in the cross-stream direction in Poiseuille flow.[12–18] More controlled motion of droplet has been observed in the presence of external effects such as electric field,[10,19–21] magnetic field,[1] temperature field,[22] acoustic wave and optical-based techniques.[1] These external fields induce interfacial stress at the fluid-fluid interface and provide a way to alter the force acting on the droplet and subsequently droplet velocity. Towards this, application of specially varying temperature field is a very effective way which alters the interfacial tension and induce a Marangoni stress at the droplet interface.[22] There is a wealth of studies in the literature which considers the sole effect of thermocapillary-induced Marangoni stress on the droplet motion in a quiescent medium. Starting from the seminal work of Young et al.,[23] several studies have considered the thermocapillary effect in the presence of fluid inertia,[24] thermal convection,[25–27] shape deformation,[28] bounding wall[29–36] and non-linear thermocapillarity effect.[37] Very recently, Choudhuri and Raja Sekhar studied the thermocapillary motion of spherical droplets in the presence of imposed background flow.[38]

Surfactants (or surface-active agents comprising of ampliphilic molecules) are integral part of droplet-based microfluidic devices.[4,39] Surfactants are used as additives in emulsification process which facilitate the generation of droplets and most importantly enhance the stability of droplets by increasing the resistance to coalesce. Hence, it is very common to have surfactants in multiphase system as additives (or sometimes as impurities also). Presence of surfactants not only reduces the interfacial tension, but also creates local gradient in interfacial tension (i.e., Marangoni stress) which has the ability to affect the motion dynamics of the droplets dramatically.[4,39] Recent studies have established a very interesting phenomenon of cross-stream migration of a spherical droplet in Poiseuille flow due to presence of surfactant-induced Marangoni stress at the fluid-fluid interface.[40–42] The non-uniformity in surfactant distribution, which creates the Marangoni stress, may be significantly altered in the combined presence of external temperature field and background Poiseuille flow.

A model which incorporates both the thermocapillary-induced and surfactant-induced Marangoni stresses at the droplet interface in the presence of background Poiseuille flow is lacking in the present literature. Towards investigating the interfacial dynamics of a surfactant-laden droplet, here, we employ both analytical and numerical techniques and obtain the droplet velocity, surfactant distribution and fluid velocity at the droplet interface. Neglecting thermal convection, fluid inertia and shape deformation, we obtain analytical solutions for the following three different asymptotic limits: (i) when the interfacial surfactant transport is dominated by the surface diffusion, (ii) when the interfacial surfactant transport is dominated by the surface convection, and (iii) when the surfactant-induced Marangoni stress is weak. Subsequently, we obtain numerical solution for wide range of governing parameters and compare with the asymptotic solutions.

## II. PROBLEM FORMULATION

### A. Physical system

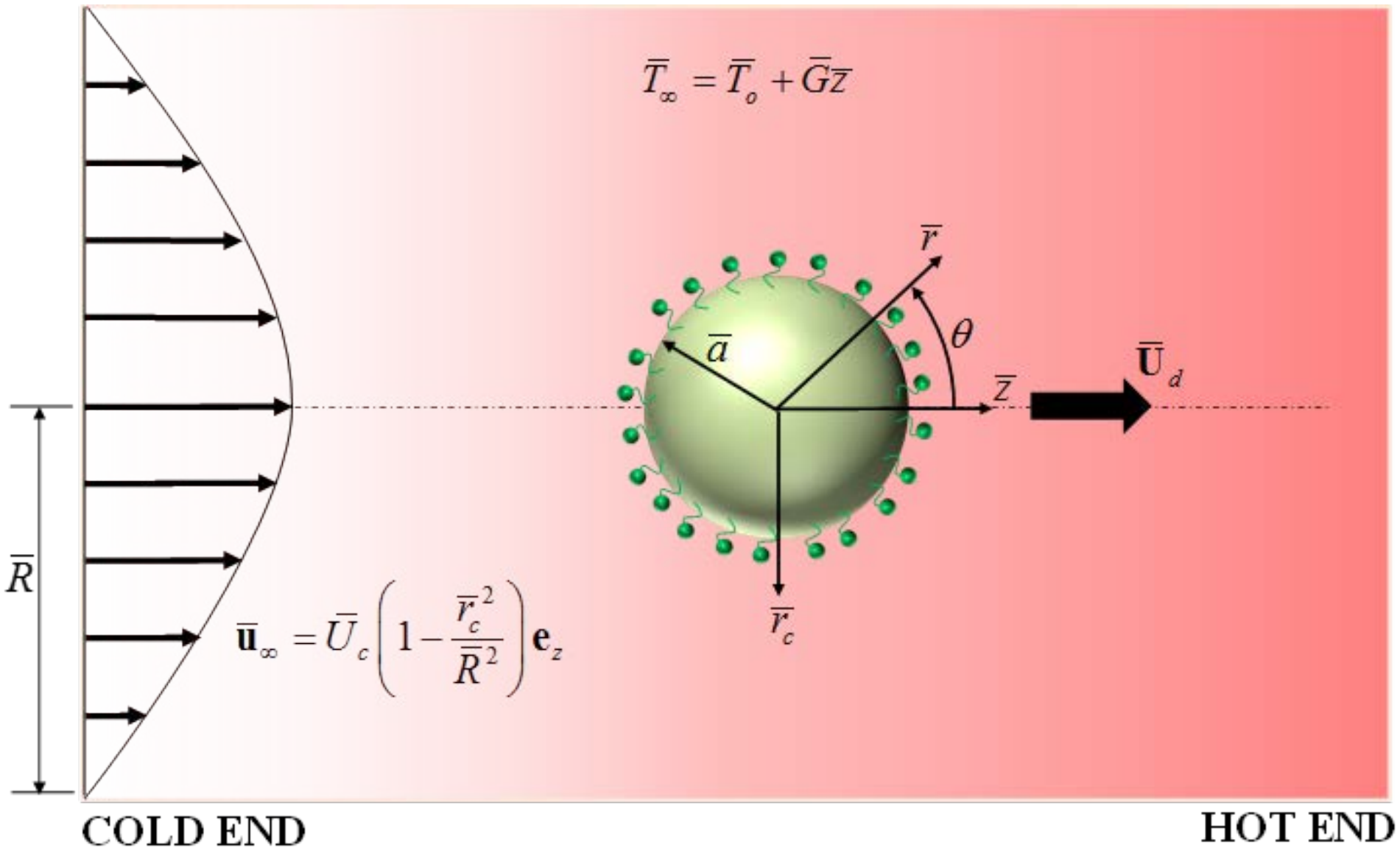

FIG. 1. Schematic representation of a surfactant-laden droplet in the presence of unbounded Poiseuille flow and linearly varying temperature field (temperature is increasing in the direction of Poiseuille flow). The droplet is spherical with radius $\bar{a}$ and moving with a velocity $\bar{\mathbf{U}}_d$. Both cylindrical $(\bar{r}_c, \bar{z})$ and spherical $(\bar{r}, \theta)$ coordinate systems are shown considering the droplet center as the origin.

The physical system under consideration consists of a Newtonian liquid droplet of radius $a$, dynamic viscosity $\mu_i$, and thermal conductivity $k_i$ suspended in another immiscible Newtonian liquid of dynamic viscosity $\mu_e$, and thermal conductivity $k_e$. Bulk-insoluble surfactants are present at the fluid-fluid interface (or droplet interface). In a quiescent medium, the surfactants are uniformly distributed over the droplet interface with a concentration of $\bar{\Gamma}_{eq}$. This surfactant-laden droplet system is acted upon by an imposed Poiseuille flow $(\bar{\mathbf{u}}_\infty)$ and an linearly varying temperature field $(\bar{T}_\infty)$. The droplet is neutrally buoyant and kept at the centerline of the Poiseuille flow (refer to Fig. 1). All the hydrodynamic and thermal properties are assumed to be constants, except the interfacial tension $\bar{\sigma}$. In the combined presence of fluid flow and temperature variation, the interfacial tension, $\bar{\sigma}\left(\bar{\Gamma},\bar{T}_s\right)$, (where $\bar{\Gamma}$ is the local surfactant concentration and $\bar{T}_s$ is the local temperature at the droplet interface) will vary at the droplet interface. Application of $\bar{\mathbf{u}}_\infty$ and $\bar{T}_\infty$ leads to motion of the droplet with a velocity $\bar{\mathbf{U}}_d = \bar{U}_d \mathbf{e}_z$. Main objective of the present study is to investigate the effect of Marangoni stress developed due to non-uniformity in temperature and surfactant distribution on $\bar{\mathbf{U}}_d$ in the presence of $\bar{\mathbf{u}}_\infty$ and $\bar{T}_\infty$. To this end, we consider an axisymmetric spherical coordinate system $(r,\theta)$ which is moving at a speed of $\bar{\mathbf{U}}_d$ and attached to the centroid of the moving droplet.

## B. Governing equations and boundary conditions

Some of the important assumptions made in the present analysis for deriving the governing differential equations and the boundary conditions are as follows: (i) The advective transport of energy is negligible as compared to its diffusive transport, which is due to very small value of the thermal Péclet number ($Pe_T = \bar{U}_c a/\alpha_e$, where $\alpha_e$ is the thermal diffusivity of suspending medium and $\bar{U}_c$ is the velocity at the channel centreline). This decouples the energy equation from the momentum equations. (ii) The convective component of acceleration is negligible, so that the flow dynamics is governed by the balance of pressure, viscous and surface tension forces. This is a valid assumption for very small value of Reynolds number $\left(Re = \rho_e \bar{U}_c a/\mu_e\right)$. (iii) We assume spherical shape of the droplet at steady state which is valid for very small capillary number $\left(Ca = \mu_e \bar{U}_c/\bar{\sigma}_o\right)$. Typical values of these non-dimensional numbers can be obtained for a methanol droplet of radius $a = 50\ \mu\text{m}$ suspended in silicone oil[43] (with $\rho_e = 955\ \text{kg/m}^3$, $\mu_e = 0.0478\ \text{Ns/m}^2$, $k_e = 0.1\ \text{W/mK}$ and $c_{pe} = 1800\ \text{J/kgK}$) as $Pe_T \sim 0.01$, $Re \sim 10^{-4}$ and $Ca \sim 0.001$ where we have taken $\bar{U}_c = 10^{-4}\ \text{m/s}$ and interfacial tension $\bar{\sigma}_o = 10^{-3}\ \text{N/m}$. With this consideration, the above three assumptions are valid in several physical situations. (iv) Surfactants are present at the droplet interface as an ideal film and does not affect the heat transfer process.[44] (v) The dependence of the interfacial tension on the surfactant concentration and temperature is taken as linear one.[45,46]

In the absence of convective transport of energy and viscous dissipation, the temperature fields inside and outside the droplet at steady state are governed by the Laplace equation of the form[44]

$$\left.\begin{aligned} \bar{\nabla}^2 \bar{T}_i &= 0, \\ \bar{\nabla}^2 \bar{T}_e &= 0, \end{aligned}\right\} \tag{1}$$

where $\bar{T}_i$ and $\bar{T}_e$ represent the temperature fields inside and outside of the droplet, respectively. Temperature fields $\left(\bar{T}_{i,e}\right)$ satisfy the following boundary conditions:[44]

(i) the temperature field outside the droplet satisfies the far-field imposed temperature: at $\bar{r} \to \infty$, $\bar{T}_e = \bar{T}_\infty = \bar{T}_o + G\bar{z}$,

(ii) $\bar{T}_i$ should be bounded inside the droplet $\left(\bar{r} < a\right)$,

(iii) temperature is continuous at the droplet interface: at $\bar{r} = a$, $\bar{T}_i = \bar{T}_e$,

(iv) heat flux is continuous at the droplet interface: at $\bar{r} = a$, $k_i \dfrac{\partial \bar{T}_i}{\partial \bar{r}} = k_e \dfrac{\partial \bar{T}_e}{\partial \bar{r}}$.

In the absence of convective transport of momentum, the velocity and pressure fields are governed by the Stokes and continuity equations of the following form[47]

$$\left.\begin{aligned} \bar{\nabla}\bar{p}_i &= \mu_i \bar{\nabla}^2 \bar{\mathbf{u}}_i,\ \bar{\nabla}\cdot\bar{\mathbf{u}}_i = 0, \\ \bar{\nabla}\bar{p}_e &= \mu_e \bar{\nabla}^2 \bar{\mathbf{u}}_e,\ \bar{\nabla}\cdot\bar{\mathbf{u}}_e = 0, \end{aligned}\right\} \tag{2}$$

where $\left(\bar{\mathbf{u}}_i, \bar{p}_i\right)$ represent the velocity and pressure fields inside the droplet, while $\left(\bar{\mathbf{u}}_e, \bar{p}_e\right)$ represent the velocity and pressure fields outside the droplet. As the flow field and temperature field are symmetric about the *z*-axis, we simplify Eq. (2) by using stream function in the following form

$$\left.\begin{aligned} \bar{\mathcal{L}}^2\left(\bar{\mathcal{L}}^2\bar{\Psi}_i\right) = 0, \\ \bar{\mathcal{L}}^2\left(\bar{\mathcal{L}}^2\bar{\Psi}_e\right) = 0, \end{aligned}\right\} \tag{3}$$

where the differential operator, $\bar{\mathcal{L}}^2$, is given by $\bar{\mathcal{L}}^2 = \dfrac{\partial^2}{\partial \bar{r}^2} + \dfrac{\left(1-\eta^2\right)}{\bar{r}^2}\dfrac{\partial^2}{\partial \eta^2}$.[47] The stream function $\left(\bar{\Psi}\right)$ is related to the velocity components in the following way

$$\bar{u}_r = -\frac{1}{\bar{r}^2}\frac{\partial \bar{\Psi}}{\partial \eta},\ \bar{u}_\theta = -\frac{1}{\bar{r}\sqrt{1-\eta^2}}\frac{\partial \bar{\Psi}}{\partial \bar{r}}, \tag{4}$$

where $\eta = \cos\theta$. The velocity and pressure fields $\left(\bar{\mathbf{u}}_{i,e}, \bar{p}_{i,e}\right)$ satisfy the boundary conditions of the following form:[44]

(i) with respect to a reference frame attached to the droplet centroid, the velocity field outside the droplet satisfies the far-field imposed velocity profile: at $\bar{r} \to \infty$, $\bar{\mathbf{u}}_e = \bar{U}_c \left(1 - \frac{\bar{r}_c^2}{\bar{R}^2}\right)\mathbf{e}_z - \bar{U}_d \mathbf{e}_z$, where $\bar{r}_c = \bar{r}\sqrt{1-\eta^2}$ is the cylindrical radial coordinate,

(ii) inside the droplet, both the velocity $(\bar{\mathbf{u}}_i)$ and pressure $(\bar{p}_i)$ fields are bounded,

(iii) at steady state, the normal components of the velocity at the droplet interface vanish: at $\bar{r} = a$, $\bar{u}_{i,r} = \bar{u}_{e,r} = 0$,

(iv) the tangential velocities at the droplet interface are continuous: at $\bar{r} = a$, $\bar{u}_{i,\theta} = \bar{u}_{e,\theta}$,

(v) the tangential hydrodynamic stress and Marangoni stress are balanced at the droplet interface: at $\bar{r} = a$, $\left[ \left(\bar{\boldsymbol{\tau}}_e - \bar{\boldsymbol{\tau}}_i\right)\cdot \mathbf{e}_r \right]\cdot \mathbf{e}_\theta = -\left(\bar{\nabla}_s \bar{\sigma}\right)\cdot \mathbf{e}_\theta$, where $\boldsymbol{\tau}_{i,e} = -\bar{p}_{i,e}\mathbf{I} + \mu_{i,e}\left[\bar{\nabla}\bar{\mathbf{u}}_{i,e} + \left(\bar{\nabla}\bar{\mathbf{u}}_{i,e}\right)^T\right]$ is the hydrodynamic stress tensor, $\mathbf{e}_r$ and $\mathbf{e}_\theta$ are the unit vectors in normal and tangential directions to spherical droplet interface, respectively. $\bar{\nabla}_s = \bar{\nabla} - \mathbf{e}_r\left(\mathbf{e}_r \cdot \bar{\nabla}\right)$ is the surface gradient operator on the spherical drop interface.

The interfacial tension, $\bar{\sigma}$, depends on the local variation of temperature and surfactant concentration at the droplet interface.[44] We assume a linear relationship of the interfacial tension with temperature and surfactant concentration in the following form:[44–46]

$$\bar{\sigma} = \bar{\sigma}_o - \beta\left(\bar{T}_s - \bar{T}_o\right) - R_g \bar{T}_o \bar{\Gamma}, \tag{5}$$

where $\bar{\sigma}_o$ is the interfacial tension at some reference temperature $T_o$, but in the absence of any surfactant. $\beta = d\bar{\sigma}/d\bar{T}$ is the gradient of interfacial tension with respect to temperature and $R_g$ is the ideal gas constant. It is to be remembered that the above linear relationship is valid only for a low concentration of surfactants.[44]

For the case of bulk-insoluble surfactants, the surfactant distribution at the droplet surface $(\bar{r} = a)$ is governed by a surface convection-diffusion equation of the form[44]

$$\bar{\nabla}_s \cdot \left(\bar{\mathbf{u}}_s \bar{\Gamma}\right) = D_s \bar{\nabla}_s^2 \bar{\Gamma}, \tag{6}$$

where $D_s$ is the surface diffusivity and $\bar{\mathbf{u}}_s = \bar{\mathbf{u}}_i\big|_{\bar{r}=a}$ is the velocity field at the droplet interface.

Now, we use the following non-dimensional scheme to obtain the relevant dimensionless parameters that govern the physical system:[44,48] $r=\bar{r}/a$, $\mathbf{u}=\bar{\mathbf{u}}/\bar{U}_c$, $p=\bar{p}\Big/\left(\dfrac{\mu_e\bar{U}_c}{a}\right)$, $\boldsymbol{\tau}=\bar{\boldsymbol{\tau}}\Big/\left(\dfrac{\mu_e\bar{U}_c}{a}\right)$, $T=\left(\bar{T}-\bar{T}_o\right)/Ga$, and $\Gamma=\bar{\Gamma}/\bar{\Gamma}_{eq}$. The non-dimensional variables are represented without overbar. Present non-dimensional scheme yields the following dimensionless property ratios:[49] viscosity ratio $\lambda=\mu_i/\mu_e$ and thermal conductivity ratio $\delta=k_i/k_e$. We also obtain the following dimensionless numbers: surface Péclet number $Pe_s=\bar{U}_c a/D_s$, thermal Marangoni number $Ma_T=\beta Ga/\mu_e\bar{U}_c$, and surfactant Marangoni number $Ma_\Gamma=\bar{\Gamma}_{eq}R\bar{T}_o/\mu_e\bar{U}_c$. The surface Péclet number signifies the relative strength of advection of surfactants as compared with diffusion at the droplet interface. The thermal Marangoni number signifies the relative strength of Marangoni stress due to non-uniform temperature distribution as compared with the viscous stress, while the surfactant Marangoni number signifies the relative strength of Marangoni stress due to non-uniform surfactant distribution as compared with the viscous stress.

Using the above scales we obtain the dimensionless version of the governing differential equation for temperature field $\left(T_{i,e}\right)$ as

$$\left.\begin{aligned}\nabla^2 T_i&=0,\\ \nabla^2 T_e&=0,\end{aligned}\right\} \tag{7}$$

with the following boundary conditions in dimensionless form

$$\left.\begin{aligned}&\text{at } r\to\infty, T_e=rP_1(\eta),\\ &T_i \text{ is bounded for } r<1,\\ &\text{at } r=1,\ T_i=T_e,\\ &\text{at } r=1, \delta\frac{\partial T_i}{\partial r}=\frac{\partial T_e}{\partial r}.\end{aligned}\right\} \tag{8}$$

The dimensionless form of the governing equations for stream function $\left(\Psi_{i,e}\right)$ is given by

$$\left.\begin{aligned}\mathcal{L}^2\left(\mathcal{L}^2\Psi_i\right)&=0,\\ \mathcal{L}^2\left(\mathcal{L}^2\Psi_e\right)&=0,\end{aligned}\right\} \tag{9}$$

subjected to the following boundary conditions in dimensionless form

$$\left.\begin{aligned}
&\text{at } r\to\infty,\ \mathbf{u}_e=\left\{1-\frac{r^2}{R^2}\left(1-\eta^2\right)-U_d\right\}\mathbf{e}_z,\\
&\mathbf{u}_i \text{ is bounded for } r<1,\\
&\text{at } r=1,\ u_{i,r}=u_{e,r}=0,\\
&\text{at } r=1,\ u_{i,\theta}=u_{e,\theta},\\
&\text{at } r=1,\ \left[\left(\boldsymbol{\tau}_e-\boldsymbol{\tau}_i\right)\cdot\mathbf{e}_r\right]\cdot\mathbf{e}_\theta=Ma_T\left(\nabla_s T_s\right)\cdot\mathbf{e}_\theta+Ma_\Gamma\left(\nabla_s\Gamma\right)\cdot\mathbf{e}_\theta.
\end{aligned}\right\} \tag{10}$$

The dimensionless form of the surfactant transport equation becomes

$$Pe_s\nabla_s\cdot\left(\mathbf{u}_s\Gamma\right)=\nabla_s^2\Gamma. \tag{11}$$

The surfactant concentration, $\Gamma$, should also satisfy the following constraint to conserve the total mass of surfactants on the droplet surface[44]

$$\int_0^\pi \Gamma\left(\theta\right)\sin\theta d\theta=2. \tag{12}$$

At this point, a very important thing to note is that the above mathematical model is non-linear due to the presence of the convective transport of surfactants at the droplet interface. The term on the left hand side of Eq. (11) is the source of non-linearity which restricts us to obtain analytical solution for any value of $Pe_s$. Another important thing to note here is that the flow field and surfactant distribution are coupled to each other. Depending on the types of surfactants, $Pe_s$ and $Ma_\Gamma$ can vary over a wide range of values. Considering $D_s=10^{-11}-10^{-8}\ \mathrm{m^2/s}$ and $\overline{\Gamma}_{eq}=10^{-10}-10^{-6}\ \mathrm{mole/m^2}$,[23,50] we obtain the ranges of $Pe_s$ and $Ma_\Gamma$ as $Pe_s=0.1-100$ and $Ma_\Gamma=0.05-500$. To solve this two-way coupled non-linear problem, we implement following two methods:[44,51] Firstly, we identify the possible asymptotic limits in the problem and use the domain perturbation method to obtain analytical solution. Secondly, we perform numerical solution of the problem for arbitrary value of $Pe_s$ and $Ma_\Gamma$ .

## III. ASYMPTOTIC SOLUTION

We obtain asymptotic solution of the present problem for the following three different limiting conditions:[44,51] (i) Small surface Péclet number limit, $Pe_s\ll 1$, which physically signifies the situation in which the convective transport of surfactant is very weak and the surfactant transport is dominated by surface diffusion. (ii) Large surface Péclet number limit, $Pe_s \gg 1$, which physically signifies the situation in which the diffusion transport of surfactant is very weak and the surfactant transport is dominated by surface convection. (iii) Small surfactant Marangoni number limit, $Ma_\Gamma \ll 1$, which physically signifies the situation in which either the effect of surfactant distribution on the variation of interfacial tension is weak or the surfactant concentration is very small. In the first two limiting conditions, $Ma_\Gamma$ can take arbitrary value, while in the third limiting condition, $Pe_s$ can take arbitrary value. In all limiting conditions, we consider $Ma_T$, $\lambda$ and $\delta$ to be arbitrary.

### A. Analytical solution for $Pe_s \ll 1$

In small surface Péclet number limit, we express any field variable $f(\mathbf{r}; Pe_s)$ in the following regular asymptotic form

$$f(\mathbf{r}; Pe_s) = f^{(0)}(\mathbf{r}) + Pe_s f^{(Pe_s)}(\mathbf{r}) + Pe_s^2 f^{(Pe_s^2)}(\mathbf{r}) + Pe_s^3 f^{(Pe_s^3)}(\mathbf{r}) + O(Pe_s^4), \qquad (13)$$

where $f^{(0)}(\mathbf{r})$ represents the leading order solution considering $Pe_s = 0$, while $f^{(Pe_s)}(\mathbf{r})$, $f^{(Pe_s^2)}(\mathbf{r})$ and $f^{(Pe_s^3)}(\mathbf{r})$ represent respective higher order correction terms which reflect effect of small $Pe_s$. Substituting the above asymptotic expansion in all the governing equations and boundary conditions, we obtain governing equations and boundary conditions which are linear at each order of perturbation. To obtain droplet velocity, which is the most important quantity of interest, we follow the following steps: As temperature field is not coupled to flow field and surfactant distribution, we first solve for temperature field. After using asymptotic expansion given in Eq. (13), we obtain that at each order of perturbation, the surfactant distribution is independent of the velocity field at that order. Hence, we solve for surfactant distribution and then go for solving flow field and obtain droplet velocity.

The leading order temperature field is governed by the Laplace equation with the far-field condition $(r \to \infty)$ as $T_e^{(0)} = rP_1(\eta)$. Solution for temperature field is classically obtained by Young, Goldstein and Block which can be adopted in the following dimensionless form

$$\left.\begin{aligned} T_i^{(0)} &= \left(\frac{3}{\delta+2}\right) r P_1(\eta), \\ T_e^{(0)} &= \left\{ r - \left(\frac{\delta-1}{\delta+2}\right)\frac{1}{r^2} \right\} P_1(\eta). \end{aligned}\right\} \qquad (14)$$

The temperature distribution at the surface of the spherical droplet is

$$T_s^{(0)} = T_i^{(0)}\Big|_{r=1} = \left(\frac{3}{\delta+2}\right)P_1(\eta). \tag{15}$$

At leading order, the surfactant distribution $\left(\Gamma^{(0)}\right)$ is governed by only diffusion transport on the droplet surface $(r=1)$ in the following form

$$\nabla_s^2\Gamma^{(0)} = 0. \tag{16}$$

Solution of Eq. (16) which satisfies the conservation of total mass of surfactant (given in Eq. (12) ) is obtained as $\Gamma^{(0)} = 1$. Hence, $\nabla_s\Gamma^{(0)} = \mathbf{0}$ and the leading order problem is simply the thermocapillary motion of droplet in Poiseuille flow. Towards solving the flow field, first we obtain the stream function distribution and the use the force-free condition to obtain droplet velocity. General solution for stream function is given by

$$\left.\begin{aligned} \Psi_i^{(0)} &= \sum_{n=1}^{\infty}\left[\hat{A}_n^{(0)}r^{n+3} + \hat{B}_n^{(0)}r^{n+1}\right]Q_n(\eta), \\ \Psi_e^{(0)} &= \Psi_\infty^{(0)} + \sum_{n=1}^{\infty}\left[C_n^{(0)}r^{2-n} + D_n^{(0)}r^{-n}\right]Q_n(\eta), \end{aligned}\right\} \tag{17}$$

where $Q_n(\eta) = \int_{-1}^{\eta} P_n(\eta)\,d\eta$ represent Gegenbauer polynomial. The stream function at far-field at leading order is given by $\Psi_\infty^{(0)} = \left(U_d^{(0)} - 1\right)r^2 Q_1 + \frac{2r^4}{5R^2}\left(Q_1 - Q_3\right)$. Using appropriate boundary conditions (given in Eq. (10) but in terms of leading order variables) and expression for surface temperature (given in Eq. (15)), we obtain the stream function distribution as

$$\left.\begin{aligned} \Psi_i^{(0)} &= \left[\hat{A}_1^{(0)}r^4 + \hat{B}_1^{(0)}r^2\right]Q_1(\eta) + \left[\hat{A}_3^{(0)}r^6 + \hat{B}_3^{(0)}r^4\right]Q_3(\eta), \\ \Psi_e^{(0)} &= \Psi_\infty^{(0)} + \left[C_1^{(0)}r + D_1^{(0)}r^{-1}\right]Q_1(\eta) + \left[C_3^{(0)}r^{-1} + D_3^{(0)}r^{-3}\right]Q_3(\eta), \end{aligned}\right\} \tag{18}$$

where expressions of $\hat{A}_n^{(0)}$, $\hat{B}_n^{(0)}$, $C_n^{(0)}$ and $D_n^{(0)}$ are given in Appendix A. Now, we use the force-free condition in the following form

$$\mathbf{F}_H^{(0)} = \mathbf{0}, \tag{19}$$

where $\mathbf{F}_H^{(0)}$ is the net hydrodynamic force acting on the droplet at leading order which is obtained as

$$\mathbf{F}_H^{(0)} = 4\pi C_1^{(0)}\mathbf{e}_z = \left\{ \frac{(3\lambda+2)\left(1-U_d^{(0)}\right)R^2 - 2\lambda}{2R^2(1+\lambda)} + \frac{Ma_T}{(\lambda+1)(\delta+2)} \right\}\mathbf{e}_z. \tag{20}$$

Substituting Eq. (20) in Eq. (19), we obtain the leading order droplet velocity as

$$U_d^{(0)} = \underbrace{1-\frac{2\lambda}{3\lambda+2}}_{\text{due to Poiseuille flow}} + \underbrace{\frac{2Ma_T}{(3\lambda+2)(\delta+2)}}_{\text{due to thermocapillary}}. \tag{21}$$

Above expression of $U_d^{(0)}$ reflects the fact that at leading order of solution the imposed Poiseuille flow and imposed temperature act independently and combined effect of these two is the linear combination obtained in Eq. (21).

With this leading order solution in hand, now, we solve for $O(Pe_s)$ problem. At $O(Pe_s)$, the temperature field is governed by the Laplace equation but temperature vanishes at far-field which gives $T_{i,e}^{(Pe_s)} = 0$ throughout the domain of solution. This is true for all higher order calculations. The surfactant transport equation at $O(Pe_s)$ is given by

$$\nabla_s^2 \Gamma^{(Pe_s)} = \nabla_s \cdot \left(\mathbf{u}_s^{(0)}\Gamma^{(0)}\right), \tag{22}$$

where the surface velocity at leading order, $\mathbf{u}_s^{(0)}$, can be obtained by using Eq. (18). Eq. (22) depicts that the $O(Pe_s)$ surfactant concentration, $\Gamma^{(Pe_s)}$, is decoupled from the $O(Pe_s)$ velocity field. We express $\Gamma^{(Pe_s)}$ in terms of Legendre polynomials in the following form

$$\Gamma^{(Pe_s)} = \sum_{n=1}^{\infty} \Gamma_n^{(Pe_s)} P_n(\eta), \tag{23}$$

where the coefficients $\Gamma_n^{(Pe_s)}$ are to be determined from Eq. (22). The left hand side of Eq. (22) can be represented as $\nabla_s^2 \Gamma^{(Pe_s)} = -\sum_{n=1}^{\infty} n(n+1)\Gamma_n^{(Pe_s)} P_n(\eta)$. By using the orthogonality of Legendre polynomial, we can obtain $\Gamma_n^{(Pe_s)}$ from the following relation

$$\Gamma_n^{(Pe_s)} = -\frac{2n+1}{2n(n+1)} \int_{-1}^{1} \nabla_s \cdot \left(\mathbf{u}_s^{(0)}\Gamma^{(0)}\right) d\eta. \tag{24}$$

Substituting $\mathbf{u}_s^{(0)}$ and $\Gamma^{(0)}$ in Eq. (24), we obtain the following non-zero coefficients

$$\left.\begin{aligned}\Gamma_1^{(Pe_s)} &= -\frac{2\delta+4+3Ma_T R^2}{(\delta+2)(2+3\lambda)R^2},\\ \Gamma_3^{(Pe_s)} &= \frac{1}{6R^2(1+\lambda)}.\end{aligned}\right\} \tag{25}$$

Using $\Gamma^{(Pe_s)} = \Gamma_1^{(Pe_s)} P_1(\eta) + \Gamma_3^{(Pe_s)} P_3(\eta)$, we obtain the stream function distribution as

$$\left.\begin{aligned}\Psi_i^{(Pe_s)} &= \left[\hat{A}_1^{(Pe_s)} r^4 + \hat{B}_1^{(Pe_s)} r^2\right] Q_1(\eta) + \left[\hat{A}_3^{(Pe_s)} r^6 + \hat{B}_3^{(Pe_s)} r^4\right] Q_3(\eta),\\ \Psi_e^{(Pe_s)} &= U_d^{(Pe_s)} r^2 Q_1 + \left[C_1^{(Pe_s)} r + D_1^{(Pe_s)} r^{-1}\right] Q_1(\eta) + \left[C_3^{(Pe_s)} r^{-1} + D_3^{(Pe_s)} r^{-3}\right] Q_3(\eta),\end{aligned}\right\} \tag{26}$$

where expressions of $\hat{A}_n^{(Pe_s)}$, $\hat{B}_n^{(Pe_s)}$, $C_n^{(Pe_s)}$ and $D_n^{(Pe_s)}$ are given in Appendix B. Similar to leading order analysis, the hydrodynamic force acting on the droplet at $O(Pe_s)$ is obtained as

$$\begin{aligned}F_H^{(Pe_s)} &= 4\pi C_1^{(Pe_s)} \mathbf{e}_z\\ &= \left[-4\pi \frac{\left\{27(2+\delta)(3\lambda+2)^2 U_d^{(Pe_s)} + 6Ma_T Ma_\Gamma\right\} R^2 + 12(2+\delta) Ma_\Gamma}{18(\lambda+1)(2+3\lambda)(2+\delta)R^2}\right] \mathbf{e}_z.\end{aligned} \tag{27}$$

Using the force-free condition, we obtain the droplet velocity at $O(Pe_s)$

$$U_d^{(Pe_s)} = -\frac{2Ma_\Gamma\left(2\delta+4+3Ma_T R^2\right)}{3(2+3\lambda)^2(\delta+2)R^2}. \tag{28}$$

The surfactant transport equation at $O\left(Pe_s^2\right)$ is given by

$$\nabla_s^2 \Gamma^{\left(Pe_s^2\right)} = \nabla_s \cdot \left(\mathbf{u}_s^{(Pe_s)} \Gamma^{(0)} + \mathbf{u}_s^{(0)} \Gamma^{(Pe_s)}\right), \tag{29}$$

where the surface velocity at $O\left(Pe_s^2\right)$, $\mathbf{u}_s^{(Pe_s)}$, can be obtained by using Eq. (26). The surfactant concentration can be decomposed in terms of Legendre polynomials as $\Gamma^{\left(Pe_s^2\right)} = \sum_{n=1}^{\infty} \Gamma_n^{\left(Pe_s^2\right)} P_n(\eta)$, where the coefficients can be determined by exploiting the orthogonality property of Legendre polynomial as

$$\Gamma_n^{\left(Pe_s^2\right)} = -\frac{2n+1}{2n(n+1)} \int_{-1}^{1} \nabla_s \cdot \left(\mathbf{u}_s^{(Pe_s)} \Gamma^{(0)} + \mathbf{u}_s^{(0)} \Gamma^{(Pe_s)}\right) d\eta. \tag{30}$$

By performing the integrations, we obtain the following non-zero coefficients

$$\Gamma^{\left(Pe_s^2\right)} = \sum_{n=1}^{4} \Gamma_n^{\left(Pe_s^2\right)} P_n\left(\eta\right) + \Gamma_6^{\left(Pe_s^2\right)} P_6\left(\eta\right), \tag{31}$$

where complete expressions of different non-zero coefficients, $\Gamma_n^{\left(Pe_s^2\right)}$, are given in Appendix C.

We obtain the stream function distribution as

$$\left.\begin{aligned} \Psi_i^{\left(Pe_s^2\right)} &= \sum_{n=1}^{4}\left[\hat{A}_n^{\left(Pe_s^2\right)} r^{n+3} + \hat{B}_n^{\left(Pe_s^2\right)} r^{n+1}\right] Q_n\left(\eta\right) + \left[\hat{A}_6^{\left(Pe_s^2\right)} r^9 + \hat{B}_6^{\left(Pe_s^2\right)} r^7\right] Q_7\left(\eta\right), \\ \Psi_e^{\left(Pe_s^2\right)} &= U_d^{\left(Pe_s^2\right)} r^2 Q_1 + \sum_{n=1}^{4}\left[C_n^{\left(Pe_s^2\right)} r^{2-n} + D_n^{\left(Pe_s^2\right)} r^{-n}\right] Q_n\left(\eta\right) + \left[C_6^{\left(Pe_s^2\right)} r^{-4} + D_6^{\left(Pe_s^2\right)} r^{-6}\right] Q_6\left(\eta\right), \end{aligned}\right\} \tag{32}$$

where expressions of $\hat{A}_n^{\left(Pe_s^2\right)}$, $\hat{B}_n^{\left(Pe_s^2\right)}$, $C_n^{\left(Pe_s^2\right)}$ and $D_n^{\left(Pe_s^2\right)}$ are not mentioned here for the sake of brevity. The hydrodynamic force acting on the droplet at $O\left(Pe_s^2\right)$ is obtained as

$$\begin{aligned} \mathbf{F}_H^{\left(Pe_s^2\right)} &= 4\pi C_1^{\left(Pe_s^2\right)} \mathbf{e}_z \\ &= \left\{-2\pi \frac{3\left(2+3\lambda\right)^3\left(\delta+2\right) R^2 U_d^{\left(Pe_s^2\right)} + 2Ma_\Gamma^2\left(2\delta+4+3Ma_T R^2\right)}{3R^2\left(\delta+2\right)\left(2+3\lambda\right)^2\left(1+\lambda\right)}\right\}\mathbf{e}_z. \end{aligned} \tag{33}$$

Using the force-free condition we obtain the droplet velocity at $O\left(Pe_s^2\right)$

$$U_d^{\left(Pe_s^2\right)} = \frac{2Ma_\Gamma^2\left(2\delta+4+3Ma_T R^2\right)}{3R^2\left(\delta+2\right)\left(2+3\lambda\right)^3}. \tag{34}$$

The surfactant transport equation at $O\left(Pe_s^3\right)$ is given by

$$\nabla_s^2 \Gamma^{\left(Pe_s^3\right)} = \nabla_s \cdot \left(\mathbf{u}_s^{\left(Pe_s^2\right)} \Gamma^{(0)} + \mathbf{u}_s^{\left(Pe_s\right)} \Gamma^{\left(Pe_s\right)} + \mathbf{u}_s^{(0)} \Gamma^{\left(Pe_s^2\right)}\right), \tag{35}$$

where the surfactant concentration can be decomposed in terms of Legendre polynomials as $\Gamma^{\left(Pe_s^3\right)} = \sum_{n=1}^{\infty} \Gamma_n^{\left(Pe_s^3\right)} P_n\left(\eta\right)$. $\Gamma_n^{\left(Pe_s^3\right)}$ can be determined by exploiting the orthogonality property of Legendre polynomial as

$$\Gamma_n^{\left(Pe_s^3\right)} = -\frac{2n+1}{2n\left(n+1\right)}\int_{-1}^{1}\nabla_s\cdot\left(\mathbf{u}_s^{\left(Pe_s^2\right)}\Gamma^{(0)} + \mathbf{u}_s^{\left(Pe_s\right)}\Gamma^{\left(Pe_s\right)} + \mathbf{u}_s^{(0)}\Gamma^{\left(Pe_s^2\right)}\right)d\eta. \tag{36}$$

By performing the integrations, we obtain the following non-zero coefficients

$$\Gamma^{\left(Pe_s^3\right)} = \sum_{n=1}^{7}\Gamma_n^{\left(Pe_s^3\right)}P_n\left(\eta\right) + \Gamma_9^{\left(Pe_s^3\right)}P_9\left(\eta\right), \tag{37}$$

where complete expressions of different non-zero coefficients, $\Gamma_n^{\left(Pe_s^2\right)}$, are not provided due to lengthy expressions.

We obtain the stream function distribution as

$$\left.\begin{aligned}\Psi_i^{\left(Pe_s^3\right)} &= \sum_{n=1}^{8}\left[\hat{A}_n^{\left(Pe_s^3\right)}r^{n+3} + \hat{B}_n^{\left(Pe_s^3\right)}r^{n+1}\right]Q_n\left(\eta\right) + \left[\hat{A}_9^{\left(Pe_s^3\right)}r^{12} + \hat{B}_9^{\left(Pe_s^3\right)}r^{10}\right]Q_9\left(\eta\right),\\ \Psi_e^{\left(Pe_s^3\right)} &= U_d^{\left(Pe_s^3\right)}r^2Q_1 + \sum_{n=1}^{8}\left[C_n^{\left(Pe_s^3\right)}r^{2-n} + D_n^{\left(Pe_s^3\right)}r^{-n}\right]Q_n\left(\eta\right) + \left[C_9^{\left(Pe_s^3\right)}r^{-7} + D_9^{\left(Pe_s^3\right)}r^{-9}\right]Q_9\left(\eta\right),\end{aligned}\right\} \tag{38}$$

where expressions of $\hat{A}_n^{\left(Pe_s^3\right)}$, $\hat{B}_n^{\left(Pe_s^3\right)}$, $C_n^{\left(Pe_s^3\right)}$ and $D_n^{\left(Pe_s^3\right)}$ are not mentioned here for the sake of brevity. Force-free conditions at $O\left(Pe_s^3\right)$ gives the droplet velocity as

$$U_d^{\left(Pe_s^3\right)} = \left\{\frac{\left(\varepsilon_1R^6Ma_T + \varepsilon_2R^4\right)Ma_\Gamma^3 + \left(\varepsilon_3R^6Ma_T^3 + \varepsilon_4R^4Ma_T^2 + \varepsilon_5R^2Ma_T + \varepsilon_6\right)Ma_\Gamma}{\varepsilon_7R^6}\right\}, \tag{39}$$

where the expressions of $\varepsilon_1 - \varepsilon_7$ are mentioned in Appendix D.

### B. Analytical solution for $Pe_s \gg 1$

In large surface Péclet number limit, we express any field variable $f\left(\mathbf{r};Pe_s^{-1}\right)$ in the following regular asymptotic form

$$f\left(\mathbf{r};Pe_s^{-1}\right) = f^{(0)}\left(\mathbf{r}\right) + Pe_s^{-1}f^{\left(Pe_s^{-1}\right)}\left(\mathbf{r}\right) + O\left(Pe_s^{-2}\right), \tag{40}$$

where $f^{(0)}\left(\mathbf{r}\right)$ represents the leading order solution considering $Pe_s \to \infty$, while $f^{\left(Pe_s^{-1}\right)}\left(\mathbf{r}\right)$ represents the $O\left(Pe_s^{-1}\right)$ correction term which reflects effect of large $Pe_s$.

The leading order temperature field is governed by the Laplace equation with the far-field condition $(r \to \infty)$ as $T_e^{(0)} = rP_1(\eta)$. So, the solution for temperature field will be exactly the same as presented in Eq. (14) and (15). At the leading order, the surfactant distribution $\left(\Gamma^{(0)}\right)$ is governed by only convective transport on the droplet surface $(r=1)$ in the following form

$$\nabla_s \cdot \left(\mathbf{u}_s^{(0)} \Gamma^{(0)}\right) = 0. \tag{41}$$

It is evident from Eq. (41) that the surfactant distribution cannot be determined from the surfactant transport equation. Hence, we solve the stream function and surfactant concentration simultaneously in the following form

$$\left.\begin{aligned} &\Psi_i^{(0)} = 0, \\ &\Psi_e^{(0)} = \Psi_\infty^{(0)} + \left[C_1^{(0)} r + D_1^{(0)} r^{-1}\right] Q_1(\eta) + \left[C_3^{(0)} r^{-1} + D_3^{(0)} r^{-3}\right] Q_3(\eta), \\ &\Gamma^{(0)} = 1 + \Gamma_1^{(0)} P_1(\eta) + \Gamma_3^{(0)} P_3(\eta), \end{aligned}\right\} \tag{42}$$

where $\Psi_\infty^{(0)} = \left(U_d^{(0)} - 1\right) r^2 Q_1 + \dfrac{2r^4}{5R^2}\left(Q_1 - Q_3\right)$ and the expression of different coefficients ($C_1^{(0)}$, $C_3^{(0)}$, $D_1^{(0)}$, $D_3^{(0)}$, $\Gamma_1^{(0)}$ and $\Gamma_3^{(0)}$) are mentioned in Appendix E. Important thing to note here is that the leading order velocity vanishes inside the droplet.

The hydrodynamic force acting on the droplet is given by

$$\mathbf{F}_H^{(0)} = 4\pi C_1^{(0)} \mathbf{e}_z = \left(\frac{3 - U_d^{(0)}}{2} - \frac{1}{R^2}\right) \mathbf{e}_z. \tag{43}$$

Substituting Eq. (20) in the force-free condition gives the leading order droplet velocity as

$$U_d^{(0)} = \left(1 - \frac{2}{3R^2}\right). \tag{44}$$

With this leading order solution in hand, now, we solve for $O\left(Pe_s^{-1}\right)$ problem. At $O\left(Pe_s^{-1}\right)$, the temperature field is governed by the Laplace equation but temperature vanishes at far-field which gives $T_{i,e}^{\left(Pe_s^{-1}\right)} = 0$ throughout the domain of solution. The surfactant transport equation at $O\left(Pe_s^{-1}\right)$ is given by

$$\nabla_s^2\Gamma^{(0)} = \boldsymbol{\nabla}_s \cdot \left( \mathbf{u}_s^{\left(Pe_s^{-1}\right)} \Gamma^{(0)} \right). \tag{45}$$

We determine the surface velocity at $O\left(Pe_s^{-1}\right)$ by using Eq. (45) in the following form

$$\begin{aligned} u_s^{\left(Pe_s^{-1}\right)} &= -\frac{\sqrt{1-\eta^2}}{\Gamma^{(0)}} \frac{d\Gamma^{(0)}}{d\eta} \\ &= \frac{3\sqrt{1-\eta^2}\left\{5(2+\delta)\left(3-7\eta^2\right)+12Ma_T R^2\right\}}{\left(35\eta^3-45\eta+12Ma_\Gamma R^2\right)(2+\delta)-36\eta Ma_T R^2}. \end{aligned} \tag{46}$$

In deriving Eq. (46) we have used the symmetry condition: $\left.\dfrac{d\Gamma^{(0)}}{d\eta}\right|_{\eta=\pm1} = 0$. At $O\left(Pe_s^{-1}\right)$ we solve stream function and surfactant distribution simultaneous in the following form

$$\left.\begin{aligned} \Psi_i^{\left(Pe_s^{-1}\right)} &= \sum_{n=1}^{\infty}\left[\hat{A}_n^{\left(Pe_s^{-1}\right)} r^{n+3} + \hat{B}_n^{\left(Pe_s^{-1}\right)} r^{n+1}\right] Q_n(\eta), \\ \Psi_e^{\left(Pe_s^{-1}\right)} &= U_d^{\left(Pe_s^{-1}\right)} r^2 Q_1 + \sum_{n=1}^{\infty}\left[C_n^{\left(Pe_s^{-1}\right)} r^{2-n} + D_n^{\left(Pe_s^{-1}\right)} r^{-n}\right] Q_n(\eta), \\ \Gamma^{\left(Pe_s^{-1}\right)} &= \sum_{n=1}^{\infty} \Gamma_n^{\left(Pe_s^{-1}\right)} P_n(\eta). \end{aligned}\right\} \tag{47}$$

where we have only determined the coefficients for $n=1$ (refer to Appendix F for detailed expression), which facilitates us to obtain the droplet velocity as

$$U_d^{\left(Pe_s^{-1}\right)} = \int_1^{-1} \frac{u_s^{\left(Pe_s^{-1}\right)} Q_1(\eta)}{\sqrt{1-\eta^2}} d\eta. \tag{48}$$

We perform numerical integration to obtain the droplet velocity using Eq. (48).

## C. Analytical solution for $Ma_\Gamma \ll 1$

In small surfactant Marangoni number limit, we express any field variable $f\left(\mathbf{r}; Ma_\Gamma\right)$ in the following regular asymptotic form

$$f\left(\mathbf{r}; Ma_\Gamma\right) = f^{(0)}(\mathbf{r}) + Ma_\Gamma f^{(Ma_\Gamma)}(\mathbf{r}) + O\left(Ma_\Gamma^2\right), \tag{49}$$

where $f^{(0)}(\mathbf{r})$ represents the leading order solution considering $Ma_\Gamma = 0$, while $f^{(Ma_\Gamma)}(\mathbf{r})$ represents the $O(Ma_\Gamma)$ correction term which reflects the effect of small $Ma_\Gamma$. The leading order temperature field is governed by the Laplace equation with the far-field condition $(r \to \infty)$ as $T_e^{(0)} = rP_1(\eta)$. So, the solution for temperature field will be exactly the same as presented in Eq. (14) and (15). At the leading order, there is no effect of surfactant on the flow field as $Ma_\Gamma = 0$. Hence, the stream function distribution and droplet velocity will be exactly the same as obtained in leading order calculation considering $Pe_s \ll 1$ (refer to Eq. (18) for stream function and Eq. (21) for droplet velocity). At leading order, the surfactant distribution $\left(\Gamma^{(0)}\right)$ is governed by convection-diffusion transport on the droplet surface $(r=1)$ in the following form

$$\begin{aligned} &Pe_s \nabla_s \cdot \left(\mathbf{u}_s^{(0)} \Gamma^{(0)}\right) = \nabla_s^2 \Gamma^{(0)}, \\ &\Rightarrow Pe_s \frac{d}{d\eta}\left\{u_s^{(0)} \Gamma^{(0)} \left(1-\eta^2\right)^{1/2}\right\} = -\frac{d}{d\eta}\left\{\frac{d\Gamma^{(0)}}{d\eta}\left(1-\eta^2\right)\right\}, \end{aligned} \tag{50}$$

where the surface velocity, $\mathbf{u}_s^{(0)} = u_s^{(0)} \mathbf{e}_\theta$, is already known. We integrate Eq. (50) and use the symmetry condition $\left.\frac{d\Gamma^{(0)}}{d\eta}\right|_{\eta=\pm 1} = 0$ and obtain the surfactant distribution at leading order as

$$\Gamma^{(0)} = k \exp(\xi), \tag{51}$$

where $\xi$ is of the following form

$$\xi = \frac{\left\{\begin{array}{l} 36 Ma_T R^2 (1+\lambda)(1-\eta) + 5(2+3\lambda)(\delta+2)\eta^3 \\ -3(10+11\lambda)(\delta+2)\eta + 2(10+9\lambda)(\delta+2) \end{array}\right\}}{12(1+\lambda)(2+3\lambda)(\delta+2)R^2} Pe_s. \tag{52}$$

We determine the term $k$ in Eq. (51) by using the total mass conservation of surfactant (refer to Eq. (12)). It is evident from Eq. (51) and (52) that though the leading order flow field and droplet velocity for $Ma_\Gamma \ll 1$ are same as the leading order solution for $Pe_s \ll 1$, the surfactant distribution is very much different.

With this leading order solution in hand, now, we solve for $O(Ma_\Gamma)$ problem. At $O(Ma_\Gamma)$, the temperature field is governed by the Laplace equation but temperature vanishes at far-field which gives $T_{i,e}^{(Ma_\Gamma)} = 0$ throughout the domain of solution. The droplet velocity at $O(Ma_\Gamma)$ can be obtained as

$$U_d^{(Ma_\Gamma)} = \frac{1}{(2+3\lambda)} \int_{-1}^{1} \Gamma^{(0)} P_1(\eta)\, d\eta. \tag{53}$$

To perform integration in Eq. (53) analytically, we approximate $\exp(\xi)$ by expanding in the following form

$$\exp(\xi) = 1 + \xi + \frac{\xi^2}{2!} + \frac{\xi^3}{3!} + \frac{\xi^4}{4!} + \frac{\xi^5}{5!} + \frac{\xi^6}{6!}, \tag{54}$$

which is found to be a good approximation for $Pe_s \sim 1$.

## IV. NUMERICAL SOLUTION

Now, we obtain droplet velocity, surfactant concentration and surface velocity for arbitrary value of $Pe_s$ and $Ma_\Gamma$ by using a numerical method. Stream function distribution is given by

$$\Psi_i = (r^2 - 1) \left[ \begin{array}{l} \left\{ \dfrac{6 + 2I_1R^2 + 3R^2(U_d - 1)}{6R^2(1+\lambda)} \right\} r^2 Q_1(\eta) + \left( \dfrac{3}{5} \dfrac{I_2}{1+\lambda} \right) r^3 Q_2(\eta) \\ + \left\{ \dfrac{6I_3R^2 - 7}{7R^2(1+\lambda)} \right\} r^4 Q_3(\eta) + \displaystyle\sum_{n=4}^{\infty} \left\{ \dfrac{n(n+1)I_n}{2(1+2n)(1+\lambda)} r^{n+1} \right\} Q_n(\eta), \end{array} \right] \tag{55}$$

$$\Psi_e = \begin{bmatrix} \left(U_d - 1\right) r^2 Q_1(\eta) + \dfrac{2}{5R^2} r^4 \left\{Q_1(\eta) - Q_3(\eta)\right\} \\ + \dfrac{1}{30R^2(1+\lambda)} \begin{pmatrix} 5\left[\left\{3(2+3\lambda)(1-U_d) + 2I_1\right\} R^2 - 6\lambda\right] r \\ + \left[\left\{15(U_d - 1)\lambda - 10I_1\right\} R^2 + 6(3\lambda - 2)\right] r^{-1} \end{pmatrix} Q_1(\eta) \\ + \left(\dfrac{3}{5}\dfrac{I_2}{1+\lambda}\right)\left(1 - r^{-2}\right) Q_2(\eta) + \dfrac{\begin{Bmatrix} \left(14 + 49\lambda + 30I_3R^2\right) r^{-1} \\ -5\left(6I_3R^2 + 7\lambda\right) r^{-3} \end{Bmatrix}}{35R^2(1+\lambda)} Q_3(\eta) \\ + \left(1 - r^{-2}\right) \sum_{n=4}^{\infty} \left[\left(\dfrac{n(n+1) I_n}{2(1+\lambda)(1+2n)}\right) r^{2-n}\right] Q_n(\eta) \end{bmatrix}. \tag{56}$$

Application of force-free condition gives the expression of droplet velocity as

$$U_d = 1 + \frac{2\left(I_1 R^2 - 3\lambda\right)}{3R^2\left(2 + 3\lambda\right)}. \tag{57}$$

In Eq. (55) - (57), only unknown is $I_n$ which can be evaluated from the following expression

$$I_n = \int_{-1}^{1} \left\{Ma_T T_s + Ma_\Gamma \Gamma\right\} P_n(\eta)\, d\eta. \tag{58}$$

The surfactant distribution $(\Gamma)$ can be determined by solving the surfactant transport equation (refer to Eq. (11)). Integrating Eq. (11) and using the symmetry condition, $\left.\dfrac{d\Gamma}{d\eta}\right|_{\eta=\pm 1} = 0$, we obtain $\Gamma(\eta)$ in terms of surface velocity as[44]

$$\Gamma(\eta) = c \exp\left[Pe_s \int_{\eta}^{1} \frac{u_s(\eta)}{\sqrt{1-\eta^2}} d\eta\right], \tag{59}$$

where $c$ is obtained from Eq. (12) as[44]

$$c = \frac{2}{\displaystyle\int_{-1}^{1} \exp\left[Pe_s \int_{\eta}^{1} \frac{u_s(t)}{\sqrt{1-t^2}} dt\right] d\eta}. \tag{60}$$

The surface velocity can be obtained from Eq. (55) in the following form

$$u_s(\eta) = -\left(\frac{1}{r\sqrt{1-\eta^2}}\frac{\partial \Psi_i}{\partial r}\right)\Bigg|_{r=1}$$

$$= -\frac{2}{\sqrt{1-\eta^2}}\left[\begin{array}{l}\left\{\frac{6+2I_1R^2+3R^2(U_d-1)}{6R^2(1+\lambda)}\right\}Q_1(\eta)+\left(\frac{3}{5}\frac{I_2}{1+\lambda}\right)Q_2(\eta) \\ +\left\{\frac{6I_3R^2-7}{7R^2(1+\lambda)}\right\}Q_3(\eta)+\sum_{n=4}^{\infty}\left\{\frac{n(n+1)I_n}{2(1+2n)(1+\lambda)}\right\}Q_n(\eta)\end{array}\right]. \tag{61}$$

A closer look into Eq. (60), (61) and (62) reveals that $\Gamma$ and $u_s$ are coupled to each other. To obtain $I_1$ (or equivalently $\Gamma_1$), we use an iterative method which is previously employed by several authors.[44,51–53] The iterative method comprises of the following steps:

(i) The droplet surface, $\eta \in [-1,1]$, is discretized in $N$ number of points. Each point is represented by $\eta_i$. We first guess the value of $\Gamma(\eta_i)$ at all the points.

(ii) To determine the surface velocity, we obtain $\Gamma_n$ by using the orthogonality of Legendre polynomial and substitute in Eq. (62). Here we truncate the infinite series upto $M$ number of terms. The choice of $M$ is based on the fact that $\Gamma_M < 10^{-4}$.

(iii) Substituting the expression of $u_s$ into Eq. (60) and (61), we obtain a new guess for $\Gamma(\eta_i)$.

(iv) With the above iterative scheme the rate of convergence is very poor. So an under-relaxation method is used to improve the convergence

$$\Gamma^{(j+1)}(\eta_i) = \beta\tilde{\Gamma}^{(j+1)}(\eta_i) + (1-\beta)\Gamma^{(j)}(\eta_i), \tag{62}$$

where $j$ is the number of iterations, $\tilde{\Gamma}^{(j+1)}(\eta_i)$ is the surfactant concentration in the $(j+1)$th iteration obtained without using the under-relaxation method and $\beta$ is the optimum relaxation parameter which varies within the range $0 \le \beta \le 1$. The choice of $\beta$ is made so that the convergence is accelerated.

(v) Above iterative scheme is executed until the following convergence criterion is satisfied

$$\max\left|\Gamma^{(j+1)}(\eta_i) - \Gamma^{(j)}(\eta_i)\right| \le 10^{-6}. \tag{63}$$

(vi) At last we use the converged $\Gamma_1$ and obtain the droplet velocity using Eq. (57).

In our numerical calculations we have used $N = 10^3$. However, the value of $M$ and $\beta$ are varied depending on $Pe_s$. For low $Pe_s$ we use $\beta = 0.5$ and $M = 10$, while for large $Pe_s$ we use $\beta = 0.005$ and $M = 40$. All the integrations involved in equations (60) and (61) are performed using the trapezoidal rule in MATLAB.

## V. RESULTS AND DISCUSSIONS

To investigate the effect of Marangoni stress generated due to non-uniform distribution of temperature and surfactants on the velocity of the droplet, we plot the variation of droplet velocity $(U_d)$ with viscosity ratio $(\lambda)$ in Fig. 2. Here we consider following two different cases: Firstly, in Fig. 2(a) we consider that the temperature at far-field is increasing in the direction of imposed Poiseuille flow (as depicted in Fig. 1). Secondly, in Fig. 2(b) and 2(c) we consider that the temperature at far-field is decreasing in the direction of imposed Poiseuille flow. Figure 2(a) depicts the droplet velocities obtained from low $Pe_s$ asymptotic solution and numerical solution. Droplet velocity solely due to imposed Poiseuille flow $(\text{i.e.}, Ma_T = Ma_\Gamma = 0)$ is shown in the inset of Fig. 2(a). It is well known that a droplet always encounters a net hydrodynamic force in the direction of hot fluid due to Marangoni stress induced by thermocapillary effect. Hence, application of temperature field increasing in the direction of imposed Poiseuille flow leads to augmentation of droplet velocity. Important thing to note here is the effect of surfactants on the droplet velocity. Figure 2(a) depicts that the presence of surfactants reduces the droplet velocity as compared with the velocity of a surfactant-free droplet. With increase in $Pe_s$, there is more reduction in droplet velocity. In this case the droplet velocity in the presence of temperature and surfactant is always more than the velocity of droplet solely due to imposed Poiseuille flow $(\text{i.e.}, Ma_T = Ma_\Gamma = 0)$. Another important thing to note here is that the effect of surfactant is more for a less viscous droplet. In the limit of $\lambda, \delta \to 0$, the droplet behaves as a bubble and encounters the effect of surfactant most significantly. This is due to the fact that in the absence of Marangoni stress, the tangential stress at the bubble interface vanishes. Hence, even a small non-uniformity in $\Gamma(\theta)$ leads to large effect on bubble velocity. In the limit of $\lambda \to \infty$, the droplet interface becomes motionless and effect of surfactant and thermocapillary vanishes. Comparison between low $Pe_s$ asymptotic solution and numerical solution reveals that the analytical solution compares very well for $Pe_s = 0.1$. However, analytical solution deviates from the numerical solution for $Pe_s = 0.2$ in the low viscosity ratio regime.

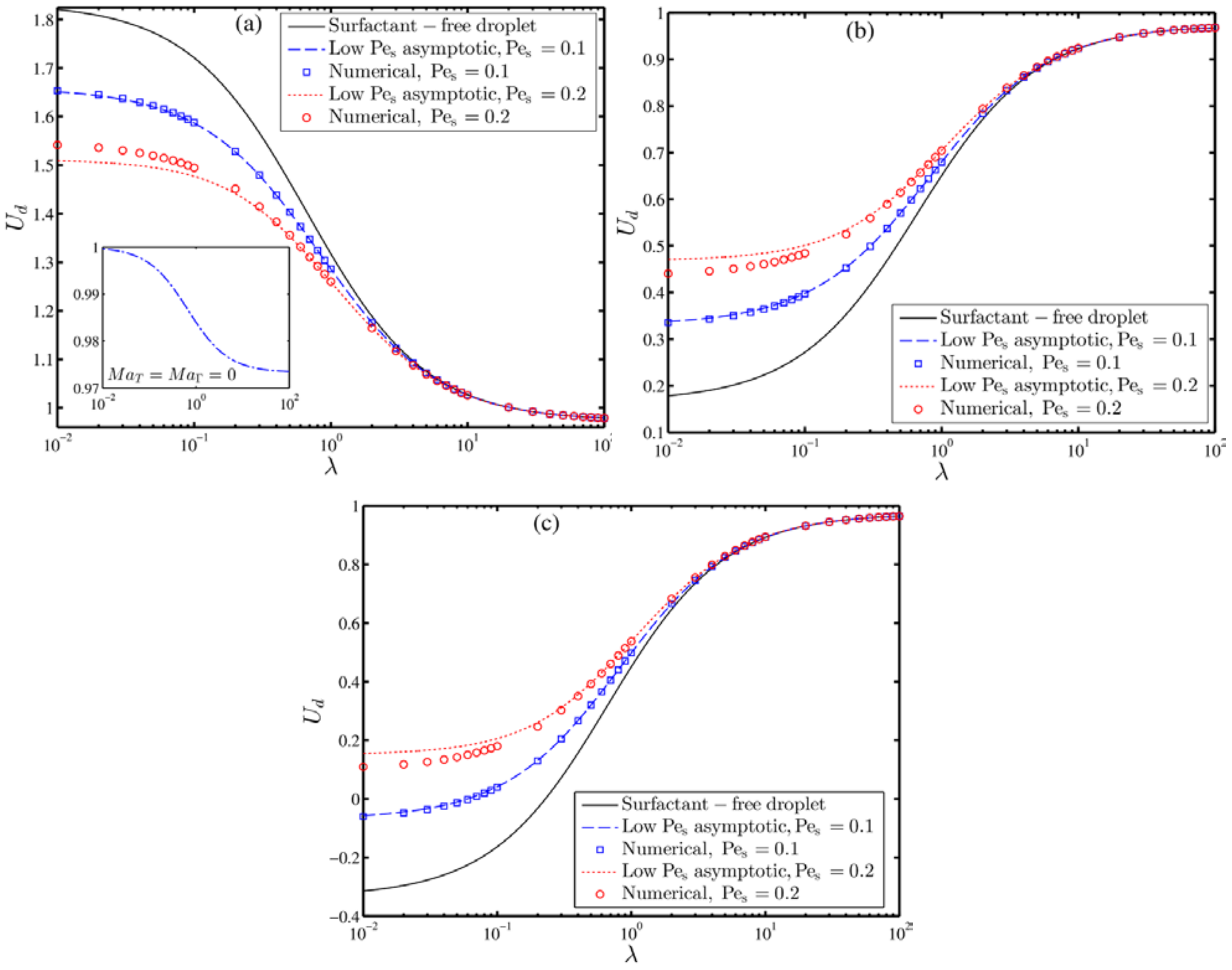

FIG. 2. Variation of droplet velocity $(U_d)$ with the viscosity ratio $(\lambda)$ for the case of (a) when temperature is increasing in the flow direction with $Ma_T = 2.5$, (b) when the temperature is decreasing in the flow direction with $Ma_T = 2.5$, and (c) when the temperature is decreasing in the flow direction with $Ma_T = 4$. Here we compare our low $Pe_s$ asymptotic solution with the numerical solutions for small values of $Pe_s$. The insets show the variation of droplet velocity in the absence of thermocapillary effect. Other parameters have the following values: $\delta = 1$ and $Ma_\Gamma = 5$.

When the far-field temperature decreases in the direction of imposed Poiseuille flow, the Marangoni stress induced due to thermocapillary effect acts to move the droplet opposite to the direction of imposed Poiseuille flow. Depending on the relative strength of imposed Poiseuille flow and thermocapillary effects (which are decided by the magnitude of $Ma_T$ and $\lambda$), droplet can move in the direction of Poiseuille flow or against it. Figure 2(b) depicts the variation of $U_d$ with $\lambda$ for $Ma_T = 2.5$. In sharp contrast to the case in which the far-field temperature increases in the direction of imposed Poiseuille flow, the droplet velocity increases in the presence of surfactants when the temperature field is reversed (refer to Fig. 2(b)). With increase in value of $Pe_s$, the droplet velocity increases. Thus, the presence of surfactants effectively negates the retarding effect of thermocapillary and the net outcome is an increase in droplet velocity. However, in this case, the droplet velocity in the presence of temperature and surfactant is always less than the velocity of droplet solely due to imposed Poiseuille flow $\left(\text{i.e.,}\ Ma_T = Ma_\Gamma = 0\right)$.

Figure 2(c) depicts the variation of $U_d$ with $\lambda$ for $Ma_T = 4$. In this case, a surfactant-free droplet moves in either direction depending on the viscosity ratio $(\lambda)$. The pivotal effect of surfactant in this case can be understood by looking into the droplet velocity for a particular value of $\lambda$. A very interesting observation to note here is that for low viscosity droplets $\left(\text{e.g.,}\ \lambda = 0.1\right)$, the non-uniformity in surfactant distribution leads to motion of the droplet in opposite direction to that of a surfactant-free droplet. Hence, the direction of droplet motion is not only governed by the direction and relative strength of imposed Poiseuille flow and temperature gradient (represented by $Ma_T$ and $\lambda$), but also decided by the strength of surfactant-induced Marangoni stress (effect of which is reflected by $Ma_\Gamma$ and $Pe_s$).

Comprehensive physical understanding about the mechanism of increase/decrease in droplet velocity due to surfactants can be obtained by investigating the interfacial flow structure, surfactant distribution and interfacial tension. First, we consider the case of increasing temperature in the direction of imposed Poiseuille flow. Figure 3(a) depicts the flow streamlines inside and outside the droplet in the presence of thermocapillary and Poiseuille flow while there is no surfactants (or the surfactants are uniformly distributed). This leads to two circulation cells inside the droplet. There are two stagnation points one at the front end $(\theta = 0)$ and other at the rear end $(\theta = \pi)$. Figure 3(b) depicts that the fluid at the droplet interface goes from front stagnation pole to rear stagnation pole. When surfactants are present at the droplet interface, this flow structure drives the surfactants away from the front stagnation pole of the droplet and surfactants are accumulated at the rear stagnation pole. Hence, the surfactant concentration reduces at $\theta = 0$ and increases at $\theta = \pi$ (refer to Fig. 3(c)). This kind of distribution of surfactants increases the local interfacial tension at the front end and decreases the same at the rear end. However, the net interfacial tension is decided by the combined effect of temperature and surfactants which is depicted in Fig. 3d. The dimensionless interfacial tension is given by $\sigma = 1 - Ca\left(Ma_T T_s + Ma_\Gamma \Gamma\right)$. In the absence of surfactants $\left(\text{i.e.,}\ Ma_\Gamma = 0\right)$, variation of $\sigma$ is solely governed by $T_s(\theta)$ (refer to the inset of Fig. 3(d)). Higher temperature near the front stagnation pole and lower temperature near the rear stagnation pole creates a gradient in the interfacial tension from front to rear pole (as depicted in Fig. 3(d)), which drives the adjacent fluid from the front to rear end and the droplet gets a net force towards the higher temperature region. This is the physical picture in the absence of surfactants.

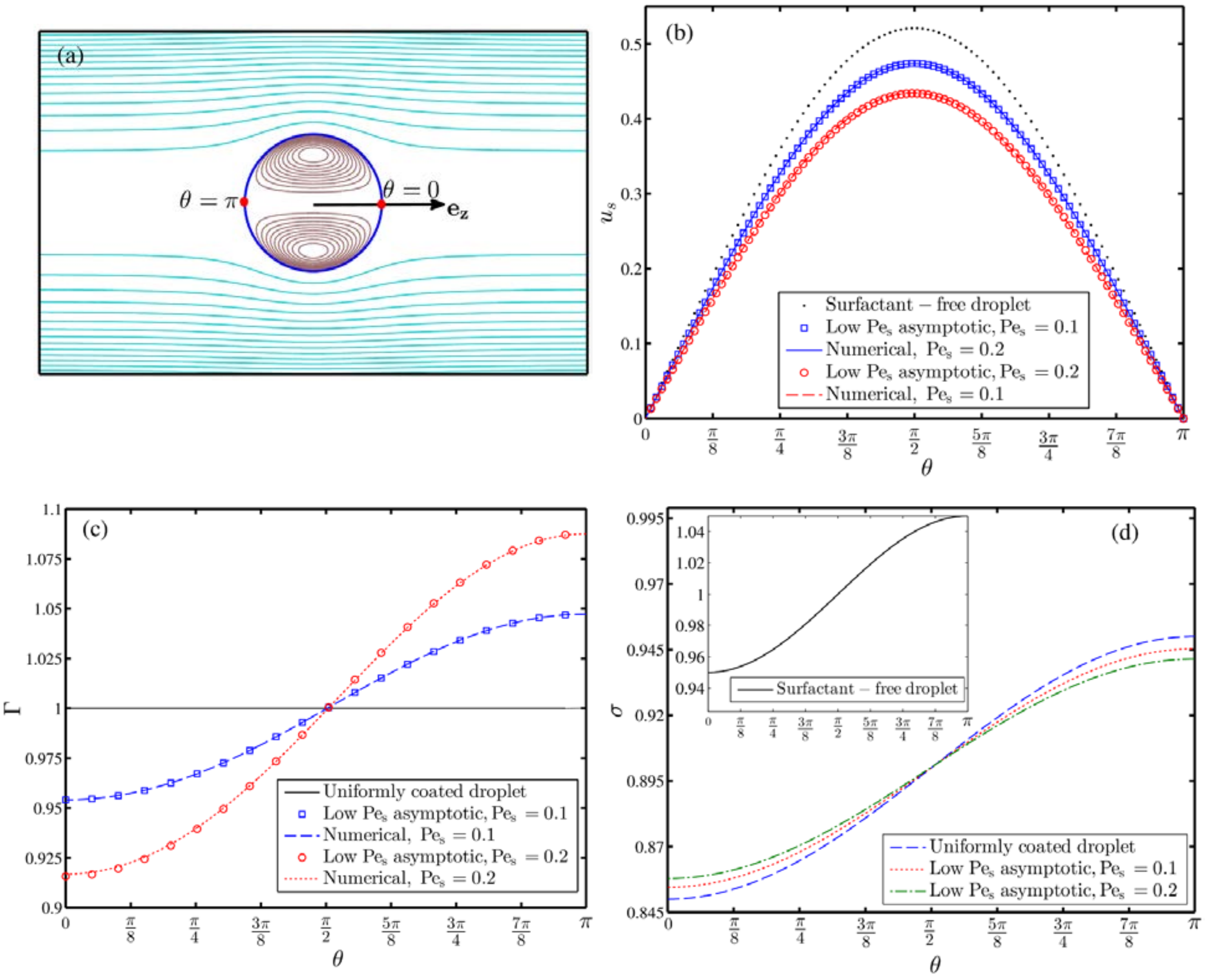

FIG. 3. (a) Streamline pattern inside and outside the droplet in the absence of surfactants, (b) Variation of surface velocity, (c) Variation of surfactant concentration, and (d) Variation of interfacial tension for the case in which the far-field temperature increases in the direction of imposed Poiseuille flow. Here we compare our low $Pe_s$ asymptotic solution with the numerical solutions for small values of $Pe_s$. Different parameters have the following values: $\delta = 1$, $\lambda = 1$, $Ma_T = 2.5$ and $Ma_\Gamma = 5$. Variation of dimensionless interfacial tension is shown considering $Ca = 0.02$.

When surfactants are present at the droplet interface, there is a decrease in interfacial tension over the droplet interface even when the surfactants are uniformly distributed $\left(\text{i.e., } Ma_\Gamma > 0, \text{ but } Pe_s = 0 \Rightarrow \Gamma(\theta) = 1\right)$. But when the surfactants are distributed non-uniformly over the droplet interface $\left(\text{i.e., } Ma_\Gamma > 0, Pe_s > 0\right)$, the front end of droplet interface encounters higher temperature but less surfactant, while the rear end encounters lower temperature but more surfactants. Thus, in this case, the surfactant-induced Marangoni stress acts opposite to the temperature-induced Marangoni stress and the net effect is a decrease in gradient in the interfacial tension. This decrease in gradient of interfacial tension leads to reduction in interfacial velocity as depicted in Fig. 3b, and subsequent retarding motion of the droplet as depicted in Fig. 2(a). Figure 3(b) and 3(c) show that the low $Pe_s$ asymptotic solution compares very well with the numerical solution for $Pe_s = 0.1$ and $0.2$.

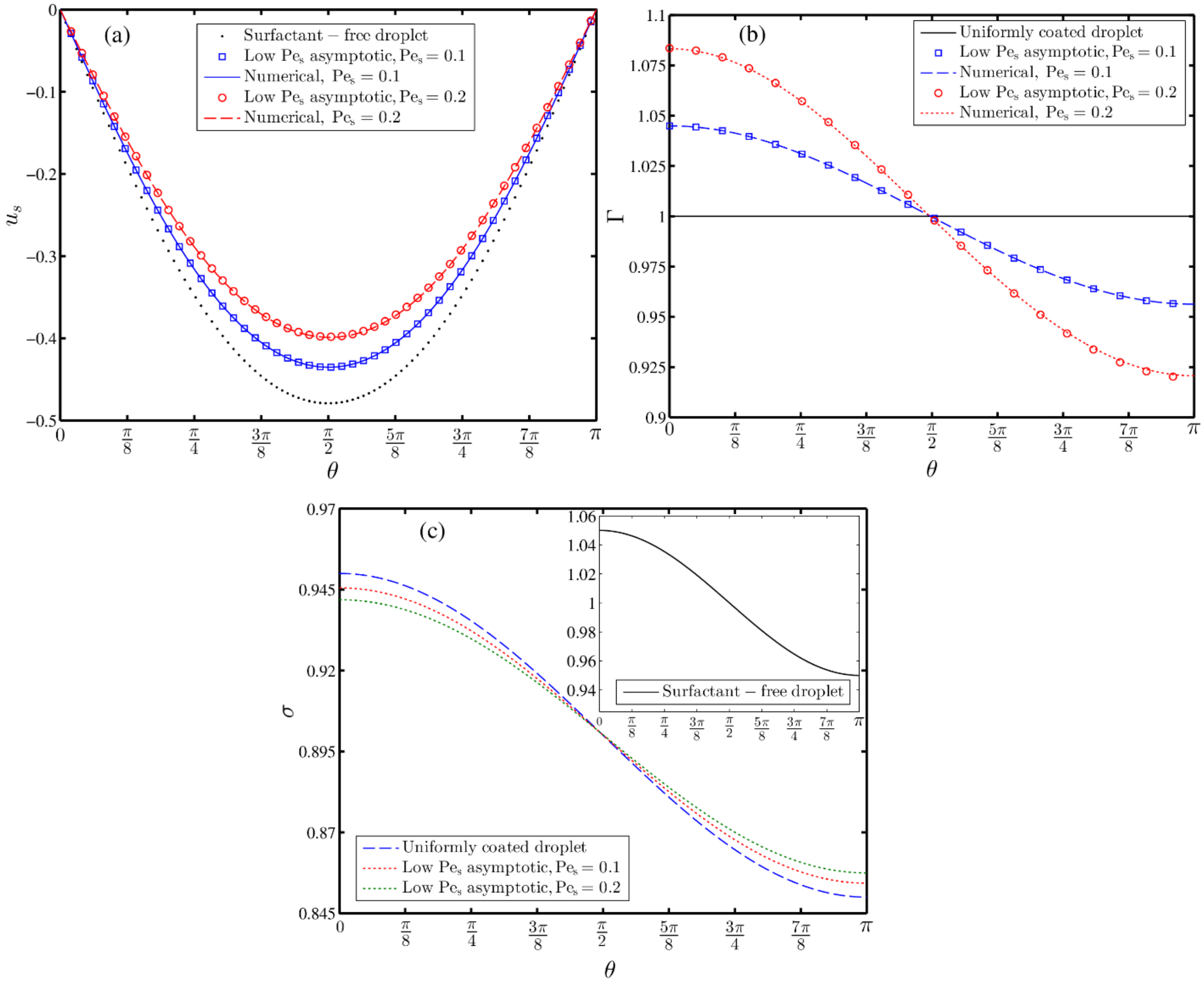

FIG. 4. (a) Variation of surface velocity, (b) Variation of surfactant concentration, and (c) Variation of interfacial tension for the case in which the far-field temperature decreases in the direction of imposed Poiseuille flow. Here we compare our low $Pe_s$ asymptotic solution with the numerical solutions for small values of $Pe_s$. Different parameters have the following values: $\delta = 1$, $\lambda = 1$, $Ma_T = 2.5$ and $Ma_\Gamma = 5$. Variation of dimensionless interfacial tension is shown considering $Ca = 0.02$.

When the far-field temperature field decreases in the direction of imposed Poiseuille flow, the magnitude of droplet velocity increases/decreases in the presence of surfactants (refer to Fig. 2(b) and 2(c)). To investigate this in more detail, we plot the surface velocity, surfactant concentration and interfacial tension in Fig. 4 for the parameters corresponding to Fig. 2(b). The flow structure (i.e., streamline pattern) remains similar as depicted in Fig. 3a. Though the droplet moves in the direction of imposed Poiseuille flow for $Ma_T = 2.5$, the surface velocity runs from rear stagnation pole $\left(\theta = \pi\right)$ to front stagnation pole $\left(\theta = 0\right)$ of the droplet (refer to Fig. 4(a)), which is due to the strong effect of thermocapillary-induced Marangoni stress at the droplet interface. When surfactants are present at the droplet interface, this surface flow drives the surfactants away from the rear stagnation pole of the droplet and surfactants are accumulated at the front stagnation pole. Figure 4(b) depicts that the surfactant concentration is less at $\theta = \pi$ and more at $\theta = 0$. So, the front end encounters lower temperature but more surfactants, and the rear end encounters higher temperature but less surfactants. Hence, the surfactant-induced Marangoni stress acts in the opposite to the temperature-induced Marangoni stress. Figure 4(c) depicts that the temperature-induced gradient in interfacial tension is reduced by the presence of surfactants. In the absence of surfactants, the gradient in interfacial tension due to thermocapillary effect retards the droplet motion when the far-field temperature decreases in the direction of Poiseuille flow (considering $Ma_T = 2.5$). So, decrease in the gradient of interfacial tension due to presence of surfactants leads to augmentation of droplet velocity (refer to Fig. 2(b)). But the surface velocity decreases (refer to Fig. 4(a)) due to the fact that the magnitude and direction of surface velocity is decided by the thermocapillary effect (for $Ma_T = 2.5$) which is now opposed by the Marangoni stress induced due to surfactants.

Now, we compare the asymptotic solutions obtained for low and high $Pe_s$ limits with the numerical solution over a wide range of $Pe_s$ in Fig. 5(a). This comparison will reveal the accuracy to which the asymptotic solutions are applicable. For the case in which the background temperature increases in the direction of imposed Poiseuille flow, with increase in $Pe_s$, numerically obtained results show a noteworthy reduction in droplet velocity. Low $Pe_s$ asymptotic solution compares well for $Pe_s < 0.3$, but starts to diverge from the numerical solution for larger values of $Pe_s$. On the other hand, the high $Pe_s$ asymptotic solution compares well with the numerical solution for $Pe_s > 50$. This establishes the fact that both the asymptotic limits are applicable in their respective limiting conditions. In the intermediate region $0.3 < Pe_s < 50$, both asymptotic theories show disagreement with the numerical solution. Now, we compare numerical solution with the third asymptotic limit of low $Ma_\Gamma$ which is valid even for $Pe_s \sim 1$ in Fig. 5(b). Figure 5(b) depicts good agreement between analytical and numerical solution for $Pe_s = 1$ but only for $Ma_\Gamma < 0.3$.

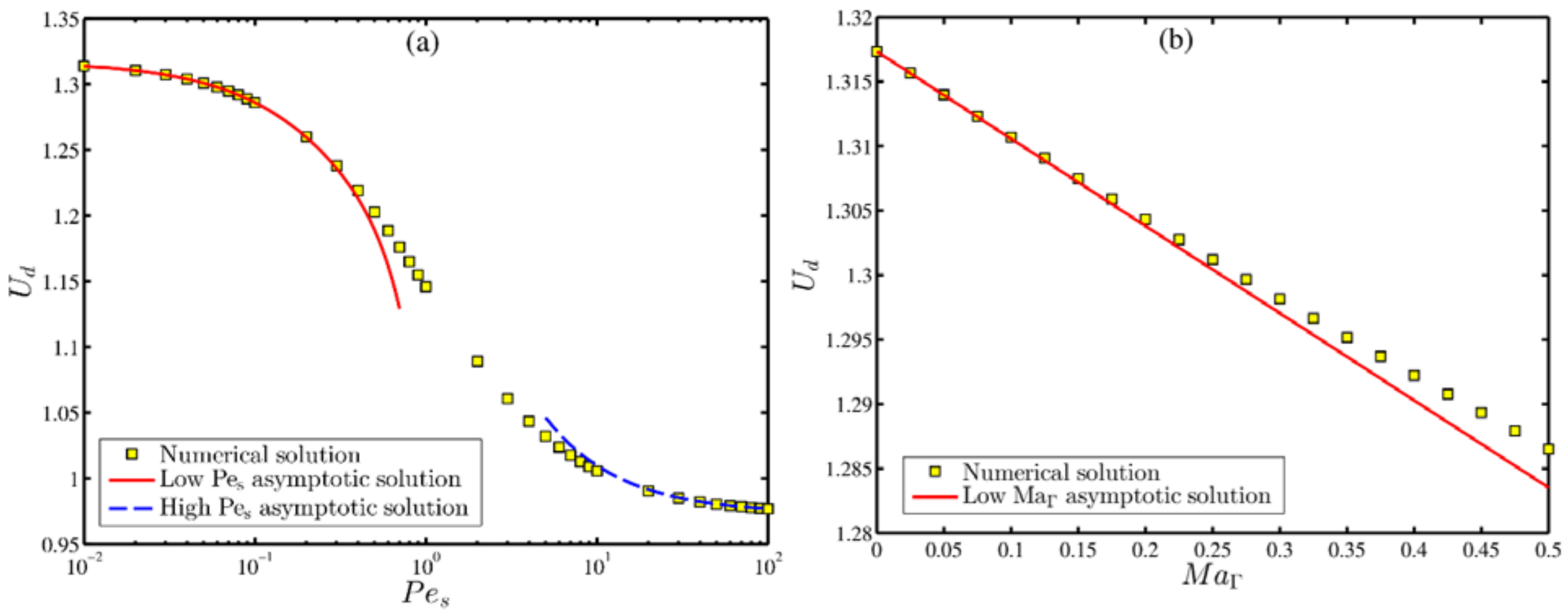

FIG. 5. (a) Comparison between numerical and analytical (low and high $Pe_s$ asymptotic) solutions for $\delta = 1$, $\lambda = 1$, $Ma_T = 2.5$ and $Ma_\Gamma = 5$. (b) Comparison between numerical and analytical (low $Ma_\Gamma$ asymptotic) solutions for $\delta = 1$, $\lambda = 1$, $Pe_s = 1$ and $Ma_\Gamma = 5$.

We show the variation of $u_s(\theta)$ and $\Gamma(\theta)$ on the droplet surface in Fig. 6(a) and 6(b) over a wide range of $Pe_s$ from numerically obtained results for the case of increasing temperature in the direction of imposed Poiseuille flow. In the absence of non-uniformity in surfactant distribution $\left(\text{i.e., } Pe_s = 0 \Rightarrow \Gamma(\theta) = 1\right)$, the fluid velocity at the droplet interface is symmetric about the equatorial place $(\theta = \pi/2)$ as depicted in Fig. 6(a). This is due to the fact that both the driving forces (i.e, thermocapillary and Poiseuille flow) independently yields symmetric velocity profile at the droplet interface. As thermocapillary and Poiseuille flow are not coupled in the absence of non-uniformity in surfactant distribution, their combined effect is just a linear combination. But for $Pe_s > 0$, thermocapillary and Poiseuille flow are coupled to each other via the surfactant transport. In this case, the interface velocity not only becomes asymmetric with respect to the equatorial plane but also reduces in magnitude. With increase in $Pe_s$, the peak in the plot of interfacial velocity is found to shift towards the front stagnation pole (refer to Fig. 6(a)). Asymmetry becomes more prominent for $Pe_s = 10$ and $50$ which is due to strong convective transport of surfactants from the front to rear end. Another important thing to note here is that the reduction in interface velocity is more near the rear end as compared with the interfacial velocity near the front end. Towards investigating this, we look into Fig. 6(b) which depicts the distribution of surfactant concentration over the droplet interface for different values of $Pe_s$. As the convective transport of surfactants takes place from the front towards the rear end, the concentration of surfactants increases significantly at the rear end. Important thing to note here is that $\Gamma(\theta)$ is also asymmetric about the $\theta = \pi/2$ which is due to the nonlinear convective transport of surfactants. It is evident form Fig. 6(b) that $|\Gamma(\theta=\pi)-1| > |\Gamma(\theta=0)-1|$. This creates a Marangoni stress which is much stronger near the rear end which finally yields more reduction in interfacial velocity near the rear end as compared to front end. We compare the high $Pe_s$ asymptotic solution with the numerical solution for $Pe_s = 100$ in the insets of Fig. 6(a) and 6(b). We plot the asymptotic solution derived for the large $Pe_s$, upto $O\left(Pe_s^{-1}\right)$. The variation of both $u_s(\theta)$ and $\Gamma(\theta)$ show good agreement between analytical and numerical solutions for $Pe_s = 100$. Effect of surfactant becomes most significant in the limit $Pe_s \to \infty$ which is the leading order solution obtained considering $Pe_s^{-1}$ as the perturbation parameter. In this limit the surfactant distribution is such that the velocity field vanishes inside the droplet and at the interface of the droplet. Droplet behaves as a spherical solid particle in this limit.

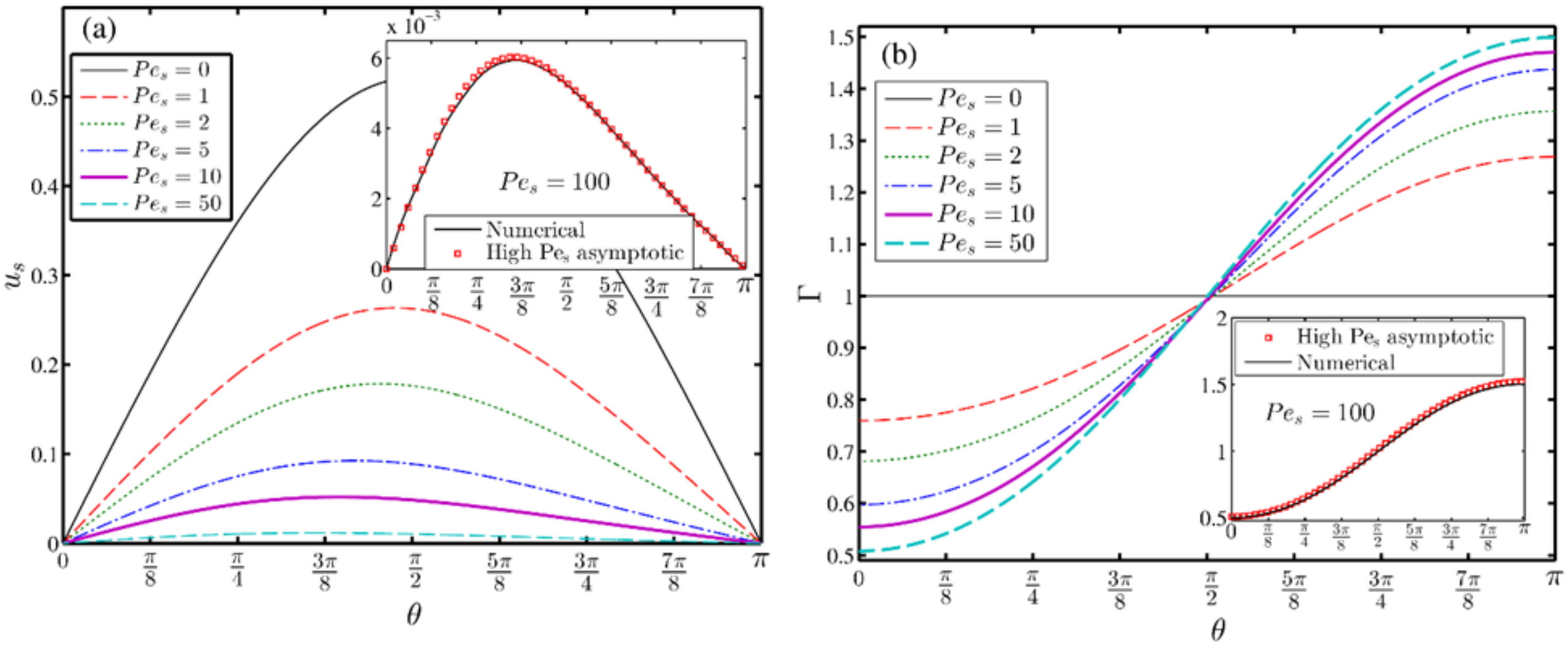

FIG. 6. (a) Variation of surface velocity. (b) Variation of surfactant distribution. The insets show the comparison between high $Pe_s$ asymptotic solution and numerical solution for $Pe_s = 100$. Different parameters are taken as $\delta = 1$, $\lambda = 1$, $Ma_T = 2.5$ and $Ma_\Gamma = 5$.

We have obtained the droplet velocity considering $Pe_s \ll 1$ in the form

$$U_d = U_d^{(0)} + Pe_s U_d^{(Pe_s)} + Pe_s^2 U_d^{\left(Pe_s^2\right)} + Pe_s^3 U_d^{\left(Pe_s^3\right)}, \tag{64}$$

where different terms are obtained in Section IIIA. Previously we have checked the validity of this equation by comparing with the numerical solution and found that the low $Pe_s$ asymptotic solution obtained upto $O\left(Pe_s^3\right)$ gives results within reasonable accuracy only for $Pe_s < 0.3$. As we have obtained a couple of terms in the asymptotic series, often it is useful to further improve the asymptotic series by using Padé approximants. The Padé approximant of order $[M/N]$ for the droplet velocity can be represented as[44,51]

$$U_d\left[M/N\right]=\frac{\sum_{n=0}^{M}a_n Pe_s^n}{1+\sum_{n=1}^{N}b_n Pe_s^n}. \tag{65}$$

So, Padé approximant represents a power series in $Pe_s$ (e.g., Eq. (65)) of degree $M+N$ in terms of the ratio of two polynomial functions in $Pe_s$ of degree $M$ and $N$. The unknown coefficients $\left(a_n \text{ and } b_n\right)$ present in the Padé approximant can be determined by equating Eq. (66) and Eq. (65) and comparing the coefficients of like powers of $Pe_s$ (starting from $Pe_s^0$ to $Pe_s^{M+N}$). The $[1/2]$ and $[2/1]$ Padé approximants are obtained as[44,51]

$$\left.\begin{aligned}U_d[1/2]&=\frac{\left(U_0^2U_2-U_0U_1^2\right)-\left(U_0^2U_3-2U_0U_1U_2+U_1^3\right)Pe_s}{\left(U_0U_2-U_1^2\right)-\left(U_0U_3-U_1U_2\right)Pe_s+\left(U_1U_3-U_2^2\right)Pe_s{}^2},\\ U_d[2/1]&=\frac{U_0U_2+\left(U_1U_2-U_0U_3\right)Pe_s+\left(U_2{}^2-U_1U_3\right)Pe_s{}^2}{U_2-U_3Pe_s},\end{aligned}\right\} \tag{66}$$

where $U_i=U_d^{\left(Pe_s^i\right)}$. To check the usefulness of the Padé approximants of droplet velocity, we plot the variation of $U_d$, $U_d\left[1/2\right]$ and $U_d\left[2/1\right]$ with $Pe_s$ along with numerical solution in Fig. 7(a). It can be seen from Fig. 6(a) that both the approximants (obtained in Eq. (67)) compare very well with the numerical solution for $Pe_s<2$. This is a noteworthy improvement as the original power series (i.e., Eq. (65)) significantly deviates from the numerical solution for $Pe_s>0.3$. We can further improve the asymptotic solution by using the Euler transformation which is often used to map singularities present on the real axis to infinity. [44,51] There are particular values of $Pe_s$ $\left(\text{e.g.}, Pe_s=\epsilon_0\right)$ for which the denominator of Eq. (67) vanishes. Considering $Pe_s=\epsilon_0$ as the singularity of the functions $U_d\left[1/2\right]$ and $U_d\left[2/1\right]$, the Euler transformation can be employed to map the singularity at infinity by constructing a new power series using the parameter $Pe_s^*=Pe_s/\left(Pe_s-\epsilon_0\right)$. Now, we can write the new power series for droplet velocity as

$$U_{b,E}=c_0+Pe_s^*c_1+Pe_s^{*2}c_2+Pe_s^{*3}c_3, \tag{67}$$

where the unknown coefficients $(c_0 - c_3)$ can be obtained by using Eq. (68) and Eq. (65) and comparing the coefficients of like powers of $Pe_s$. For $U_d[2/1]$ we obtain $\epsilon_0 = U_2/U_3$ and droplet velocity after taking Euler transformation is obtained as[44,51]

$$U_{d,E}[2/1] = U_0 - U_1\epsilon_0 Pe_s^* + \left(U_2\epsilon_0^2 - U_1\epsilon_0\right)Pe_s^{*2} - \left(U_3\epsilon_0^3 - 2U_2\epsilon_0^2 + U_1\epsilon_0\right)Pe_s^{*3}. \qquad (68)$$

Now, we compare $U_{d,E}[2/1]$ with the numerical simulation in Fig. 7(b). Figure 7(b) shows that there is a remarkable improvement of the power series and it compares very well over a wide range of $Pe_s$. Another Euler transformation can also be obtained from $U_d[1/2]$ which gives very similar results to that of $U_{d,E}[2/1]$.

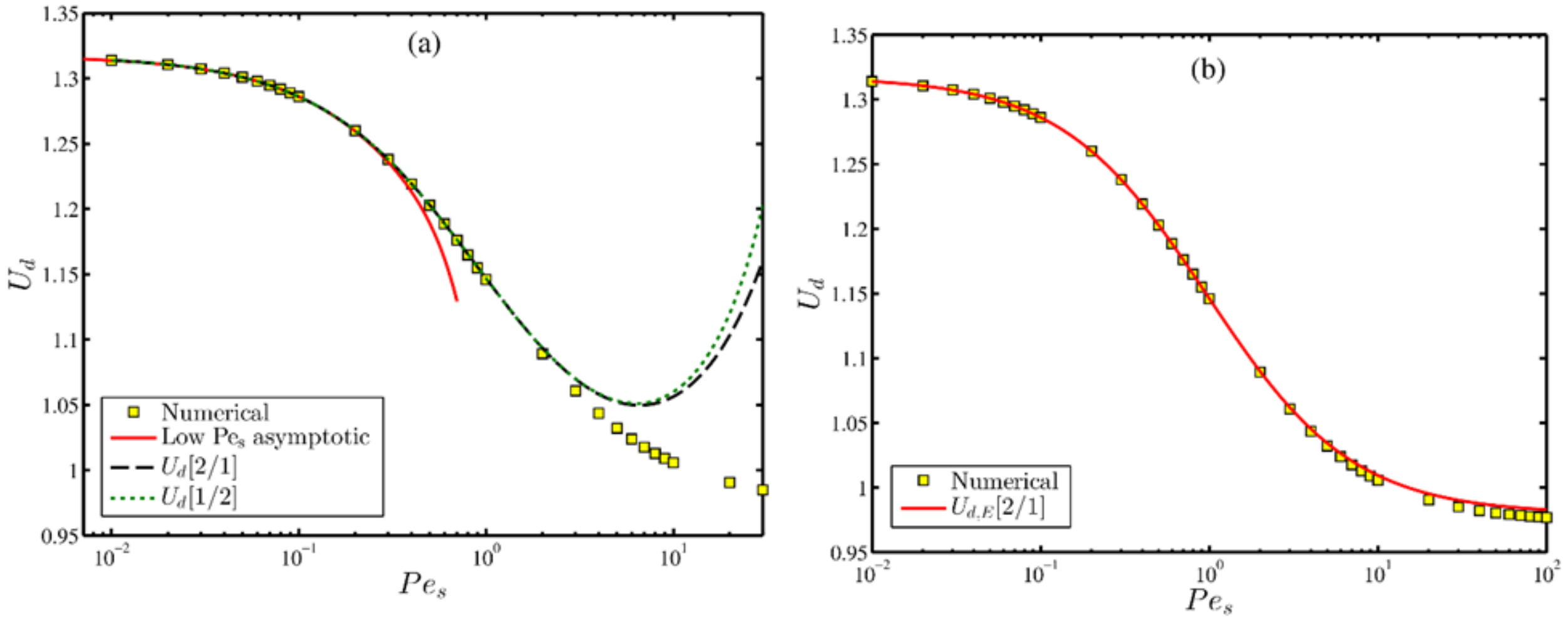

FIG. 7. (a) Comparison between numerical solution and analytical solutions (low $Pe_s$ asymptotic and two Padé approximants). (b) Comparison between numerical solution and low $Pe_s$ asymptotic solution after taking Euler transformation. Different parameters are taken as $\delta = 1$, $\lambda = 1$, $Ma_T = 2.5$ and $Ma_\Gamma = 5$.

## VI. CONCLUSIONS

Droplet motion in Poiseuille flow in the presence of Marangoni stress is analyzed assuming thermal convection, fluid inertia and shape deformation to be negligible. In the present model, the Marangoni stress includes both the effect of temperature and bulk-insoluble surfactants. The Marangoni stress induced due to non-uniform surfactant distribution is controlled by $Pe_s$ and $Ma_\Gamma$. We obtain asymptotic solution for the following three different limits: (i) $Pe_s \ll 1$, (ii) $Pe_s \gg 1$ and (iii) $Ma_\Gamma \ll 1$. We have employed a numerical scheme to obtain solution for arbitrary values of $Pe_s$ and $Ma_\Gamma$. The present study shows that in the absence of non-uniformity in surfactant distribution, the droplet velocity due to imposed Poiseuille flow and due to thermocapillary effect can be linearly combined to obtain the net effect. In this case the magnitude and direction of droplet motion are governed by the direction of imposed Poiseuille flow and temperature gradient and relative magnitudes of $Ma_T$ and $\lambda$. However, in the presence of non-uniformity in surfactant distribution, this linear combination is not applicable which is due to the non-linear and coupled nature of the convective transport of surfactants at the droplet interface. When the externally applied temperature increases in the direction of the imposed Poiseuille flow, the droplet motion is retarded by the surfactant-induced Marangoni stress. Interesting things are obtained for the case in which the externally applied temperature decreases in the direction of imposed Poiseuille flow. In this case, the droplet motion may be augmented or retarded depending on the magnitude of $Ma_T$, $Ma_\Gamma$, $\lambda$ and $Pe_s$. For particular values of these parameters, a surfactant-laden droplet moves in opposite direction to a surfactant-free droplet. When the advective transport of surfactant is strong (i.e., large value of $Pe_s$), we observe significant reduction in fluid velocity at the droplet interface and asymmetric distribution of surface velocity surfactant concentration. Asymptotic solutions compare well with the numerical solution but only in respective limiting conditions. Use of the Padé approximants and Euler transformation further improved the low $Pe_s$ asymptotic series which compares very well with numerical solution over a wide range of $Pe_s$.

## Appendix A: Expression of the constant coefficients of stream function for leading order

The expressions of the constant coefficients present in the leading order stream functions (refer to Eq. (18)) are obtained as

$$\left.\begin{aligned}
&\hat{A}_1^{(0)}=\frac{1}{2}\frac{\left\{\left(U_d^{(0)}-1\right)\delta+2Ma_T+2U_d^{(0)}-2\right\}R^2+4+2\delta}{\left(2+\delta\right)R^2\left(\lambda+1\right)},\ \hat{B}_1^{(0)}=-\hat{A}_1^{(0)},\\
&\hat{A}_3^{(0)}=-\frac{1}{R^2\left(1+\lambda\right)},\ \hat{B}_3^{(0)}=-\hat{A}_3^{(0)},\\
&C_1^{(0)}=\frac{\left\{\begin{array}{l}2\left(1-U_d{}^{(0)}\right)\delta-4U_d{}^{(0)}+2\left(2+Ma_T\right)\\-3\left(U_d{}^{(0)}-1\right)\left(2+\delta\right)\lambda\end{array}\right\}R^2-2\lambda\left(2+\delta\right)}{2\left(2+\delta\right)\left(\lambda+1\right)R^2},C_3^{(0)}=\frac{2+7\lambda}{5R^2\left(1+\lambda\right)},\\
&D_1^{(0)}=\frac{\left\{5\left(U_d{}^{(0)}-1\right)\left(2+\delta\right)\lambda+10Ma_T\right\}R^2+2\left(3\lambda-2\right)\left(2+\delta\right)}{10\left(2+\delta\right)R^2\left(\lambda+1\right)},\ D_3^{(0)}=-\frac{\lambda}{R^2\left(1+\lambda\right)}.
\end{aligned}\right\}\quad(69)$$

## Appendix B: Expression of the constant coefficients of stream function for $O\left(Pe_s\right)$

The expressions of the constant coefficients present in the leading order stream functions (refer to Eq. (26)) are obtained as

$$\left.\begin{aligned}
&\hat{A}_1^{(Pe_s)}=\frac{\left[\left\{\left[3\left(2+3\lambda\right)\left(2+\delta\right)\right]U_d{}^{(Pe_s)}-6Ma_TMa_\Gamma\right\}R^2-4\left(\delta+2\right)Ma_\Gamma\right]}{6R^2\left(\lambda+1\right)\left(2+3\lambda\right)\left(2+\delta\right)},\\
&\hat{B}_1^{(Pe_s)}=-\hat{A}_1^{(Pe_s)},\hat{A}_3^{(Pe_s)}=\frac{Ma_\Gamma}{R^2\left(1+\lambda\right)\left(7+7\lambda\right)},\ \hat{B}_3^{(Pe_s)}=-\hat{A}_3^{(Pe_s)},\\
&C_1^{(Pe_s)}=-\frac{\left\{9\left(2+\delta\right)\left(3\lambda+3\right)^2U_d{}^{(Pe_s)}+6Ma_TMa_\Gamma\right\}R^2+4\left(2+\delta\right)Ma_\Gamma}{6\left(\lambda+1\right)\left(2+3\lambda\right)\left(2+\delta\right)R^2},\\
&D_1^{(Pe_s)}=\frac{\left\{3\left(2+\delta\right)\left(3\lambda+2\right)\lambda U_d{}^{(Pe_s)}+6Ma_TMa_\Gamma\right\}R^2+4\left(2+\delta\right)Ma_\Gamma}{6\left(\lambda+1\right)\left(2+3\lambda\right)\left(2+\delta\right)R^2},\\
&C_3^{(Pe_s)}=\frac{Ma_\Gamma}{R^2\left(\lambda+1\right)\left(7+7\lambda\right)},\ D_3^{(Pe_s)}=-\frac{Ma_\Gamma}{R^2\left(\lambda+1\right)\left(7+7\lambda\right)}.
\end{aligned}\right\}\quad(70)$$

## Appendix C: Expression of the constant coefficients of surfactant concentration for $O\left(Pe_s{}^2\right)$

The expressions of the constant coefficients present in the $O\left(Pe_s^2\right)$ surfactant concentration (refer to Eq. (31)) are obtained as

$$
\left.\begin{aligned}
&\Gamma_1^{\left(Pe_s^2\right)} = \frac{Ma_\Gamma\left(2\delta+4+3Ma_T R^2\right)}{R^2\left(\delta+2\right)\left(2+3\lambda\right)^2},\ \Gamma_3^{\left(Pe_s^2\right)} = -\frac{1}{42}\frac{Ma_\Gamma}{R^2\left(1+\lambda\right)^2},\ \Gamma_6^{\left(Pe_s^2\right)} = \frac{25}{4158\left(1+\lambda\right)^2 R^4},\\
&\Gamma_2^{\left(Pe_s^2\right)} = \frac{1}{3}\frac{\left\{\begin{aligned}&\left(\frac{1}{2}\delta+1+Ma_T R^2\right)\lambda\\ &+\left(\frac{5}{9}\delta+\frac{10}{9}+Ma_T R^2\right)\end{aligned}\right\}\left\{\begin{aligned}&\left(\frac{13}{21}\delta+\frac{26}{21}+Ma_T R^2\right)\lambda\\ &+\left(\frac{40}{63}\delta+\frac{80}{63}+Ma_T R^2\right)\end{aligned}\right\}}{R^4\left(\delta+2\right)^2\left(\frac{2}{3}+\lambda\right)^2\left(1+\lambda\right)^2},\\
&\Gamma_4^{\left(Pe_s^2\right)} = -\frac{1}{27720}\frac{\left(2640Ma_T R^2+1670\delta+3340\right)\lambda+1700\delta+3400+2640Ma_T R^2}{R^4\left(1+\lambda\right)^2\left(\delta+2\right)\left(\frac{2}{3}+\lambda\right)}.
\end{aligned}\right\} \quad (71)
$$

## Appendix D: Expression of the constants present in droplet migration velocity of $O\left(Pe_s{}^3\right)$

The expressions of the constant coefficients present in the $O\left(Pe_s^2\right)$ droplet velocity (refer to Eq. (39)) are obtained as

$$
\left.\begin{aligned}
&\varepsilon_1 = -124740\left(1+\lambda\right)^3\left(\delta+2\right)^2,\\
&\varepsilon_2 = -83160\left(1+\lambda\right)^3\left(\delta+2\right)^3,\\
&\varepsilon_3 = 74844(1+\lambda)^3,\\
&\varepsilon_4 = 5346\left(31\lambda+30\right)\left(1+\lambda\right)^2\left(\delta+2\right),\\
&\varepsilon_5 = 132\left(1+\lambda\right)\left(909\lambda^2+1770\lambda+860\right)\left(\delta+2\right)^2,\\
&\varepsilon_6 = \left(28593\lambda^3+26600+81860\lambda+83850\lambda^2\right)\left(\delta+2\right)^3,\\
&\varepsilon_7 = 62370\left(1+\lambda\right)^3\left(\delta+2\right)^3\left(2+3\lambda\right)^4.
\end{aligned}\right\} \quad (72)
$$

## Appendix E: Expression of the constant coefficients of stream function and surfactant concentration in Eq. (42)

The expressions of the constant coefficients present in the $O\left(Pe_s^2\right)$ stream function and surfactant concentration (refer to Eq. (42)) are obtained as

$$\left.\begin{aligned}&C_1^{(0)}=-\frac{3R^2\left(U_d^{(0)}-1\right)+2}{2R^2},\ D_1^{(0)}=\frac{6+5R^2\left(U_d^{(0)}-1\right)}{10R^2},\ C_3^{(0)}=\frac{7}{5R^2},\\&D_3^{(0)}=-\frac{1}{R^2},\ \Gamma_1^{(0)}=-\frac{4+2\delta+3Ma_T R^2}{Ma_\Gamma R^2\left(2+\delta\right)},\Gamma_3^{(0)}=\frac{7}{6Ma_\Gamma R^2}.\end{aligned}\right\}\tag{73}$$

## Appendix F: Expression of the constant coefficients of stream function and surfactant concentration for $O\left(Pe_s^{-1}\right)$

The coefficients present in $O\left(Pe_s^{-1}\right)$ stream function and surfactant concentration for $n=1$ (refer to Eq. (47)) are obtained as

$$\left.\begin{aligned}&\hat{A}_1^{\left(Pe_s^{-1}\right)}=\frac{1}{2}F_1,\ B_1^{\left(Pe_s^{-1}\right)}=-\frac{1}{2}F_1,\ C_1^{\left(Pe_s^{-1}\right)}=-\frac{3}{2}U_d^{\left(Pe_s^{-1}\right)}+\frac{1}{2}F_1,\\&D_1^{\left(Pe_s^{-1}\right)}=\frac{1}{2}U_d^{\left(Pe_s^{-1}\right)}-\frac{1}{2}F_1,\ \Gamma_1^{\left(Pe_s^{-1}\right)}=\frac{1}{2}\frac{F_1\left(3\lambda+2\right)}{Ma_\Gamma},\end{aligned}\right\}\tag{74}$$

where $F_1=3\int_{-1}^{1}u_s^{\left(Pe_s^{-1}\right)}Q_1\left(\eta\right)d\eta.$

## ACKNOWLEDGMENT

The authors would like to acknowledge Prof. R Shankar Subramanian (Department of Chemical Engineering, Clarkson University) for sending one chapter from the PhD thesis of HS Kim.

The authors gratefully acknowledge the financial support provided by the Indian Institute of Technology Kharagpur, India [Sanction Letter no.: IIT/SRIC/ATDC/CEM/2013-14/118, dated 19.12.2013].